\documentclass[final,5p,times,twocolumn]{elsarticle}
\usepackage{times,amsmath,epsfig}
\usepackage{subfigure}
\usepackage{booktabs}
\usepackage{hyperref}
\usepackage{array}
\usepackage{graphicx}
\usepackage{epstopdf}
\usepackage{textcomp}
\usepackage{booktabs}               
\usepackage{diagbox}                  
\usepackage{multirow}

\usepackage{perpage}
\MakePerPage{footnote}

\newcommand{\eat}[1]{}

\newcommand{\tabincell}[2]{\begin{tabular}{@{}#1@{}}#2\end{tabular}}

\journal{Journal of Neurocomputing}

\begin{document}

\begin{frontmatter}

\title{Deep Learning in Computer-Aided Diagnosis and Treatment of Tumors: A Survey}

\author[hebut1,hebut2]{Dan Zhao}
%\ead{201831404018@hebut.stu.edu.cn}
\author[hebut1,hebut2]{Guizhi Xu\corref{cor1}}
\cortext[cor1]{Co-first author}
\author[hebut1,hebut2]{Zhenghua Xu\corref{cor2}}
\ead{zhenghua.xu@hebut.edu.cn}
\cortext[cor2]{Corresponding author}
\author[oxford]{Thomas Lukasiewicz}
\author[983]{Minmin Xue}
\author[983]{Zhigang Fu}
%\ead{thomas.lukasiewicz@cs.ox.ac.uk}
%\cortext[cor2]{Principal corresponding author}
%\fntext[fn1]{This is the specimen author footnote.}
%\fntext[fn2]{Another author footnote, but a little more longer.}
\address[hebut1]{State Key Laboratory of Reliability and Intelligence of Electrical Equipment, Hebei University of Technology, China}
\address[hebut2]{Key Laboratory of Electromagnetic Field and Electrical Apparatus Reliability of Hebei Province, Hebei University of Technology, China}
%\address[bupt1]{Beijing Key Laboratory of Intelligent Telecommunications Software and Multimedia, Beijing University of Posts and Telecommunications, China}
\address[oxford]{Department of Computer Science, University of Oxford, UK} 
\address[983]{Department of Health Management Center, 983 Hospital of Joint Logistics Support Force, Tianjin, 300142, China}





\begin{abstract}
Computer-Aided Diagnosis and Treatment of Tumors is a hot topic of deep learning in recent years, which constitutes a series of medical tasks, such as detection of tumor markers, the outline of tumor leisures, subtypes and stages of tumors, prediction of therapeutic effect, and drug development. Meanwhile, there are some deep learning models with precise positioning and excellent performance produced in mainstream task scenarios. Thus follow to introduce deep learning methods from task-orient, mainly focus on the improvements for medical tasks. Then to summarize the recent progress in four stages of tumor diagnosis and treatment, which named In-Vitro Diagnosis (IVD), Imaging Diagnosis (ID), Pathological Diagnosis (PD), and Treatment Planning (TP). According to the specific data types and medical tasks of each stage, we present the applications of deep learning in the Computer-Aided Diagnosis and Treatment of Tumors and analyzing the excellent works therein. This survey concludes by discussing research issues and suggesting challenges for future improvement.
\end{abstract}


\end{frontmatter}

%\linenumbers

\section{Introduction}
% 1 page
\label{sec:Introduction}

Cancer is a leading cause of death and a notable public health problem worldwide. According to Global Cancer Statistics~\cite{bray2018global}, by 2018, there will be an estimation of 18.1 million new cancer cases and 9.6 million deaths caused by cancer. Thus to diagnose early and treat compatibly becomes one of the most important research topics in medicine. Previously, people devote to the simple medical method to approach the above goal, which depends on the information obtained by the doctor's perception and the experiences gained by years. While the perception can be influenced by objective factors (e.g., sensory threshold, fatigue, and prior knowledge.) and the acquisition of experience often costs more time, and such experience is often subjective. With the tremendous development of artificial intelligence, considerable attention has been paid to plentiful applications based on artificial intelligence that was used for Computer-Aided Diagnosis and Treatment of Tumors because this technique could assist solve the above two issues. Since Hinton et al.~\cite{hinton2006reducing} prove that traditional statistical methods are not necessary to extract features as long as computing resources are sufficient, deep neural networks have been asked to learn useful latent features from big data. After that, deep learning has become a new force in Computer-Aided Diagnosis and Treatment of Tumors.  

Recently, deep learning differentiates many delicate branches to perform different tasks for different data types. The two most commonly branches are Computer Vision (CV), Natural Language Processing (NLP). In Computer Vision, deep learning works well in many tasks, such as classification, object detection, semantic segmentation. Also, in Natural Language Processing, these tasks (such as text classification, question answering, text abstraction) could accomplish using deep learning. For different requirements of diverse tasks, there are generating several state-of-the-art models. Commonly, these models have their specialties and focus on the specific work.
For instance, Convolutional Neural Network (CNN~\cite{lecun1998gradient}) works very well on classification; Region-CNN series (R-CNNs~\cite{girshick2014rich, girshick2015fast, ren2015faster}) are excellent on object detection; Fully Convolutional Network (FCN~\cite{long2015fully}) and U-Net~\cite{ronneberger2015u} do well in semantic segmentation; Recurrent Neural Network (RNN~\cite{lipton2015critical}) and Long Short-Term Memory (LSTM~\cite{hochreiter1997long}) are good at text classification. 

Due to the ability of data fusion in deep learning and the needs to model the data features of tumor in diagnosis and treatment process, there are numerous researches to construct Computer-Aided Diagnosis and Treatment of Tumors based on deep learning. Therefore, many recent efforts have been conducted to summarize the relevant researches in this area~\cite{ker2017deep, meyer2018survey, yasaka2018deep, sahiner2019deep, hu2018deep, ueda2019technical, liu2018applications, mazurowski2019deep, liu2019deep, napel2018quantitative, cao2018deep, shen2017deep, razzak2018deep, ching2018opportunities, akkus2019survey}.

Mainly, Shen et al. summarized the contributions in computer-aided analysis of medical images, such as image registration, structural detection, tissue segmentation, computer-aided disease diagnosis, and prognosis~\cite{shen2017deep}. Ching et al. introduced the opportunities and obstacles to deep learning regarding biology, which focuses on the perspective of medicine~\cite{ching2018opportunities}. Sahiner et al. focus on Convolutional Neural Networks, investigated image segmentation, detection, characterization, processing and reconstruction, tasks involving imaging and treatment in medical imaging and radiotherapy, and discussed methods for dataset expansion~\cite{sahiner2019deep}.

However, there exist some drawbacks to these review works. In brief, like the work from shen et al.~\cite{shen2017deep}, these works based on respective medical tasks~\cite{cao2018deep, razzak2018deep, ching2018opportunities}, which contents , lack systematization. Similar to Ching et al.~\cite{ching2018opportunities} focus on biology and medicine, these works based on a single type of medical imaging~\cite{liu2018applications, mazurowski2019deep, liu2019deep, napel2018quantitative, akkus2019survey}, ignoring that the diagnosis and treatment of tumor is essentially a process of multi-department coordination and multi-technology integration in clinical practice. Also, like the work from Sahiner et al.~\cite{sahiner2019deep}, these researchers whose work based on a few deep learning models and does not clearly explain the relationship and advantages of each state-of-the-art method in deep learning~\cite{ker2017deep, meyer2018survey, yasaka2018deep, hu2018deep, ueda2019technical}. Furthermore, some review works are kinds of outdated~\cite{shen2017deep, ker2017deep}.

Therefore, in this work, we aim to summarize the recent progress (basically concentrated in 2017-2019) in Computer-Aided Diagnosis and Treatment of Tumors to fill the above gaps. The contributions of this work were shown as follow:
\begin{itemize}
\item This paper reviews the medical applications by deep learning in 2017-2019, which introduces neoteric state-of-the-art and latest applications.
\item This paper introduces the primary network of deep learning from task-oriented and expounds the superior model of improving network according to the characteristics of medical tasks.
\item This paper follows a clear clue - the process of Computer-Aided Diagnosis and Treatment of Tumors, which organized a transparent knowledge system to summarize recent work.
\item This paper will service clinical and computer researchers, which improvement can optimize the knowledge structure for both of them and promote interdisciplinary mutual understanding and learning for each other.
\item Given the existing gaps in Computer-Aided Diagnosis and Treatment of Tumors, this paper looks forward to the challenges in data types, organs, medical tasks, deep learning methods. Also, this paper points out the blank in adversarial study like the work of Finlayson et al.~\cite{finlayson2019adversarial}, which needs to be paid attention by researchers.
\end{itemize}

\section{Classic Deep Learning Models and Improvements}
\label{sec:Deeplearningmodels}

This section reviews classic deep learning models which widely used in Computer-Aided Diagnosis and Treatment of Tumors. Thus to mainly introduce Convolutional Neural Network (CNN~\cite{lecun1998gradient}), Fully Convolutional Network (FCN~\cite{long2015fully}), Region-CNN (R-CNN~\cite{girshick2014rich}) and their improved models.

It is worth mentioning that there are still exist state-of-the-art in-depth learning model series work well on computer science, such as Mask Region-CNN (Mask R-CNN~\cite{he2017mask}) in instance segmentation and Transformer~\cite{vaswani2017attention} series in Natural Language Processing (NLP). However, in Computer-Aided Diagnosis and Treatment of Tumors, the target organ, lesion, or tissue of tumors are generally stable in structure. It means that the instance segmentation hard to play to its strengths at the current research stage, which is especially good at multi-class and multi-object detection and segmentation. Also, clinical data (such as clinical case, prescription, and treatment plan) has never received the attention it deserves, which could be intellectualized by NLP. Therefore, this paper chooses to introduce CNN, FCN, and RCNN series in detail.

\subsection{Convolutional Neural Network and DeepMedic}

\subsubsection{CNN}

\begin{figure}[!t]
\begin{center}
      \epsfig{figure=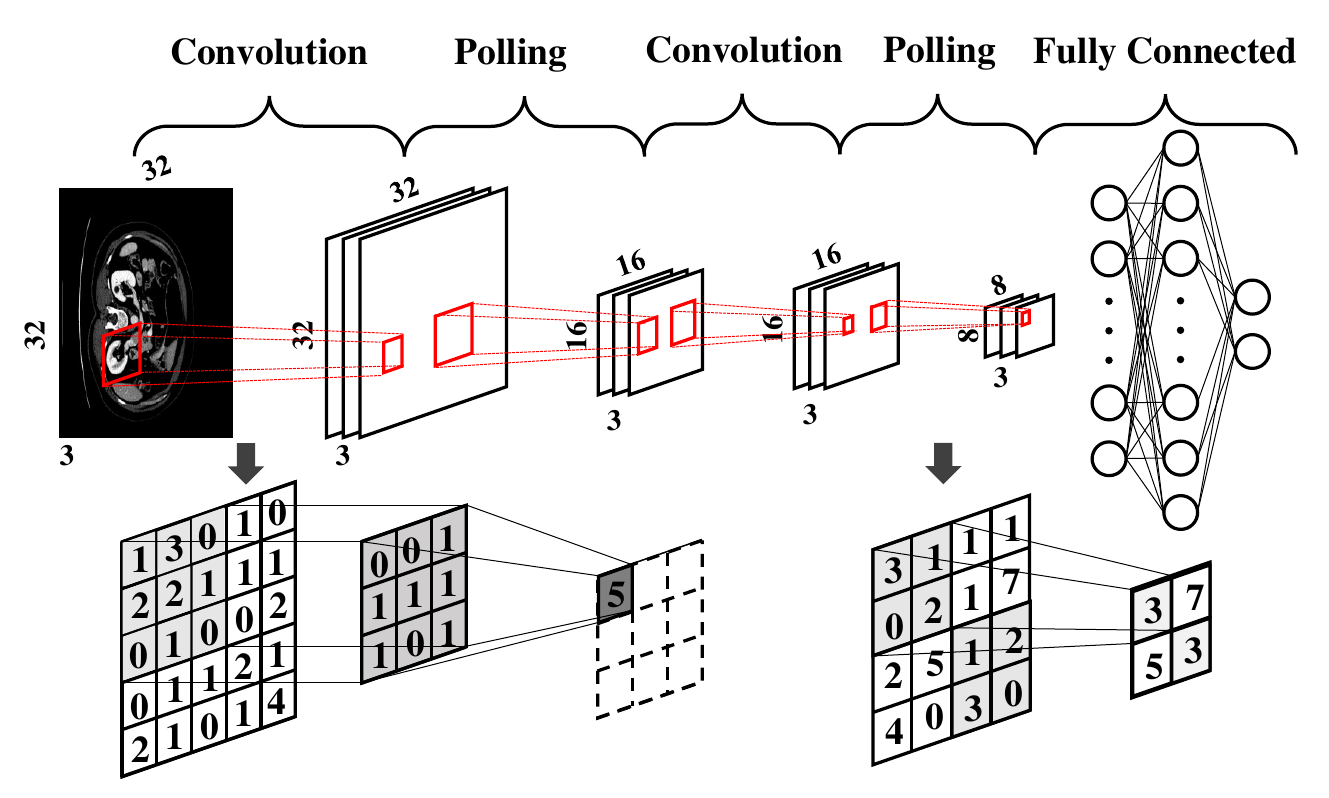,angle=0, width=3.25in}
\end{center}
\caption{The structure of CNN.}%\vspace{-3mm}
\label{cnn}
\end{figure}

Convolutional Neural Network (CNN~\cite{lecun1998gradient}) was widely used in medical tasks as the foundational deep learning method. The structure of CNN with details shows in Figure~\ref{cnn}. There are three integral parts named the convolutional layer, the pooling layer, and the Fully Connected layer (FC). Followed by the pooling layer, there will be an activate function that could enhance the ability to fit nonlinear problems.

Stem from CNN's unique structure - the convolutional layer realizes the local region connection and weight sharing, the pooling layer achieves dimension reduction, and the fully connected layer implements the task of the classifier. There are four ways to improve the CNN except for exchange activation function.

\begin{enumerate}
\item Specific to the fully connected layer, researchers exchange this layer into other networks such as Conditional Random Fields (CRF~\cite{kamnitsas2017efficient}). This improvement makes CNN no longer limited to classification tasks but keeps the ability of the feature extraction.
\item Increasing the depth of CNN by stacking multiple convolution and pooling layers. However, this method hard to achieve a super deep network due to the problems of gradient vanishing and gradient exploding. Therefore, Visual Geometry Group Network (VGGnet~\cite{simonyan2014very}) adopted 16-layer and 19-layer structures.
\item To solve the degradation problem with skip connection, such as Deep Residual Network (ResNet~\cite{he2016deep}), solves the problem of the higher error rate generated in a deeper network.
\item Increasing the diversity of features by adding convolution kernel of different sizes and introducing the dimension reduction of 1*1 convolution kernel, such as GoogLeNet~\cite{szegedy2015going}.
\end{enumerate}

Finally, Table~\ref{tab1} shown the recent works in Computer-Aided Diagnosis and Treatment of Tumors using CNN.

\begin{table}[htbp]
\centering
\caption{CNN in Computer-Aided Diagnosis and Treatment of Tumors}
\label{tab1}
\setlength{\tabcolsep}{3mm}{
\begin{tabular}{cc}
\toprule
{Tasks} & {Reference}\\
\midrule
{Detection} & {\cite{bellver2017detection}}\\
{Location} & {\cite{milletari2017hough}}\\
{Classification} & {\cite{araujo2017classification}, \cite{cruz2017accurate}, \cite{saltz2018spatial}, \cite{golatkar2018classification}, \cite{khoshdeli2018deep}, \cite{khosravi2018deep}}\\
{Segmentation} & \tabincell{c}{\cite{saouli2018fully}, \cite{li2017deep}, \cite{milletari2017hough}, \cite{laukamp2019fully}, 
\cite{leung2018deep}, \cite{trebeschi2017deep},\\ \cite{li2018tumor}, \cite{janowczyk2018resolution}, \cite{bellver2017detection}}\\
{Prediction} & \tabincell{c}{\cite{li2017deep}, \cite{jackson2018deep}, \cite{foote2018real}, \cite{zhen2017deep}, \cite{oakden2017precision}, \cite{zhang2017personalized},\\ \cite{li2017deep}, \cite{bychkov2018deep}, \cite{wang2019deep}, \cite{zhao20183d}}\\
\bottomrule
\end{tabular}}
\end{table}

The location and detection focus on pointing out the position of the object, except the latter one could distinguish different objects. Table~\ref{tab1} shows that CNN has been used to detect or locate the tumor in medical tasks, which usually is the first step of a two-stage work, such as locating or detecting first and then segmenting for the MRI of brain~\cite{milletari2017hough}. Similarly, Bellver et al. exchange the last layer into a single neuron to detect healthy/unhealthy liver tissues~\cite{bellver2017detection}. However, these researches using CNN are often ancillary products of tasks because the detection or location was not the ultimate purpose of the research. Thus detection and location using CNN cannot be regarded as independent medical tasks. There is another computer science method named Region-CNN series (R-CNNs), which will introduce after, are the start-of-the-art of object detection.

The classification is CNN's turf, thanks to its excellent ability of feature extraction, lots of researchers prefer to address the medical data by CNN. As the Table \ref{tab1} shows, these researchers focus the tumor diagnosis and treatment on breast~\cite{araujo2017classification, cruz2017accurate, golatkar2018classification, khosravi2018deep}, kidney~\cite{khoshdeli2018deep}, bladder~\cite{khosravi2018deep}, and even 13 cancer types~\cite{saltz2018spatial} based on histology images.

Segmentation is another application of CNN which was demanded a lot in medical tasks, weather the initial screening, subtype identification or treatment of tumors all revolve around the precise segmentation. Researchers in this area usual adopt the first variant (exchange the last layer in CNN) we introduce above to achieve the goal of lesion's segmentation. In details, the researches in Table~\ref{tab1} reflects the extensive applications of segmentation using CNN to deal with multi-modal data and multi-tissue organs, such as brain's MRI~\cite{saouli2018fully, li2017deep, milletari2017hough, laukamp2019fully}, rectum's MRI~\cite{trebeschi2017deep}, nasopharynx's MRI~\cite{li2018tumor}, lung's PET~\cite{leung2018deep}, liver's CT~\cite{bellver2017detection}, prostate's~\cite{bulten2019epithelium} and breast's~\cite{janowczyk2018resolution} histology image.

The prediction can be regarded as an extension of classification, since its essence is to classify whether or not a certain situation will occur in the future. In general, we usually need to make the following predictions, such as tumor growth prediction~\cite{zhang2017personalized}, tumor invasiveness prediction~\cite{zhao20183d}, motion estimation of radiotherapy targets~\cite{foote2018real}, radiation dose estimation~\cite{jackson2018deep}, radiation toxicity prediction~\cite{zhen2017deep}, outcome prediction~\cite{bychkov2018deep}, recurrence prediction~\cite{wang2019deep}, and life prediction~\cite{oakden2017precision}.

Above all, CNN is a method with excellent feature extraction capability and is specialized in classification tasks. These two advantages of CNN depend on its structure. Based on this, researchers began to improve the model from four aspects that we reviewed before to promote the performance of CNN, make it easy to train and adapt to specific tasks. In tumor diagnosis and treatment, CNN has used to initially screen tumors, determine subtypes, outline tumors, and even predict treatment plans and outcomes. All in all, CNN can be said to be one of the most widely used models of Computer-Aided Diagnosis and Treatment of Tumors.

\subsubsection{DeepMedic}

\begin{figure}[!t]
\begin{center}
      \epsfig{figure=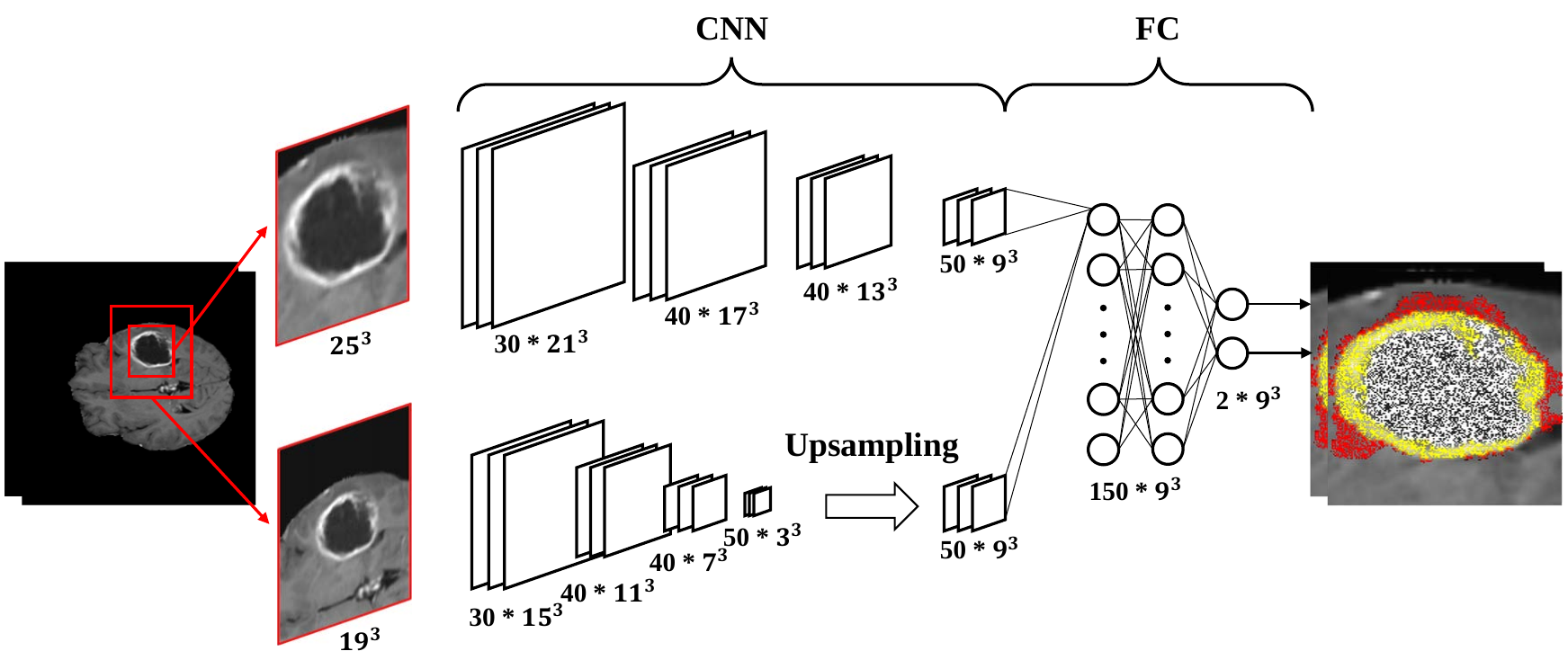,angle=0, width=3.25in}
\end{center}
\caption{The structure of DeepMedic.}%\vspace{-3mm}
\label{DeepMedic}
\end{figure}

DeepMedic is an efficient 11-layers deep, multi-scale, 3D CNN architecture, which is an improved model specific to brain MRI images from BRATS 2015 and ISLES 2015~\cite{kamnitsas2017efficient}.
Figure~\ref{DeepMedic} is the structure of DeepMedic. The baseline of DeepMedic is a CNN, followed by Conditional Random Fields (CRF).

It proposes a solution with parallel convolutional pathways for multi-scale processing, which efficiently incorporate both local and contextual information and improves the segmentation results. At last, the author uses CRF to achieve space regularization.

The most valuable contribution of DeepMedic is dense-training, which can balance the distribution of training samples from different segmentation classes, which is a massive problem in medical datasets because, in clinical practice, there are many tiny objects. Meanwhile, DeepMedic could make a dense prediction over multiple adjacent pixels at one time, thus saving computing costs. In recent years, there are several works based on DeepMedic, which shows in Table~\ref{tab2}.

\begin{table}[htbp]
\centering
\caption{DeepMedic in Computer-Aided Diagnosis and Treatment of Tumors}
\label{tab2}
\setlength{\tabcolsep}{9mm}{
\begin{tabular}{cc}
\toprule
{Tasks} & {Reference}\\
\midrule
{Segmentation} & {\cite{perkuhn2018clinical}, \cite{kao2018brain}, \cite{kamnitsas2017ensembles}, \cite{castillo2017volumetric}, \cite{laukamp2019fully}}\\
{Prediction} & {\cite{kao2018brain}}\\
\bottomrule
\end{tabular}}
\end{table}

DeepMedic is a trial variant of CNN that it focuses on the segmentation of medical tasks. Also, because of the original use for brain MRI, the most common application scenario is still the segmentation of brain MRI, which shows in Table~\ref{tab2}. The majority of the researches in the table is segmentation due to the improvement in the last layer. This structure makes the DeepMedic better in the brain's segmentation. Except for the work of Kamnitsas et al.~\cite{kao2018brain}, they predict the population survival using SVM after the specific lesion was segmented.

In general, DeepMedic's improvements to medical data are groundbreaking and widely used in brain tumor segmentation tasks, but its bright spot is also its shortcoming. That is, its applications are limited to brain segmentation. In this respect, u-net, which will be introduced later, breaks the stereotype that the improvement of the deep learning model based on medical data is just an appendage of deep learning and will not affect the development in the field of computer science.

\subsection{Fully Convolutional Network and U-Net}

\subsubsection{FCN}

\begin{figure}[!t]
\begin{center}
      \epsfig{figure=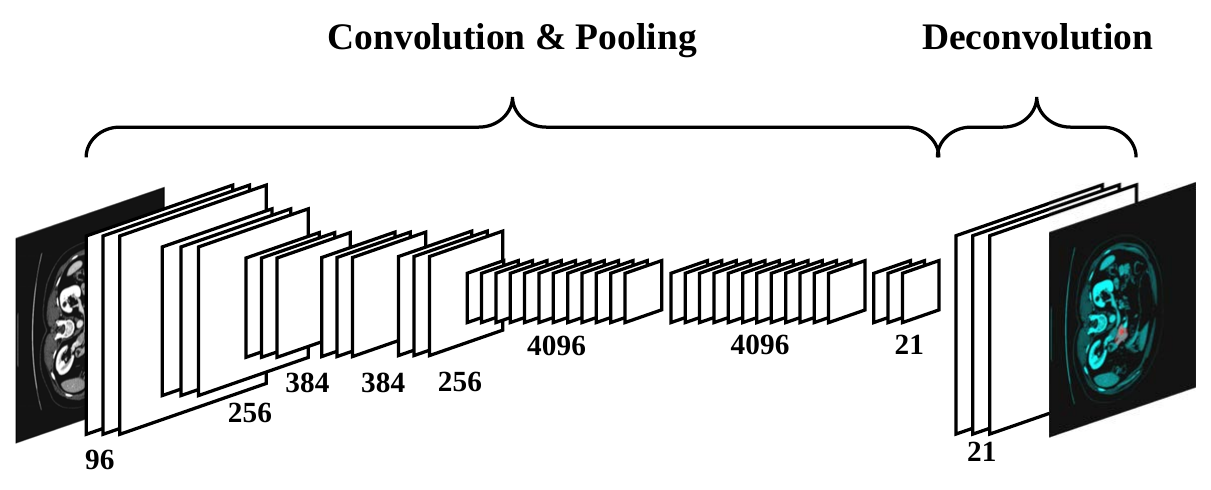,angle=0, width=3.25in}
\end{center}
\caption{The structure of FCN.}%\vspace{-3mm}
\label{fcn}
\end{figure}

As an improved method, Fully Convolutional Network (FCN~\cite{long2015fully}) exchanged the fully connected layer into a deconvolution layer. The structure of the FCN is shown in Figure~\ref{fcn}, which is an encoder-decoder model. The form of FCN provides a foundation for the future development of semantic segmentation. In this way, FCN allows end-to-end pixel-level classification, the original size of the feature map in up-sampling, and several roughnesses of the up-sampling. Above all, FCN can accept input images of any size and do well in medical image segmentation at medical tasks.

Because of the structure, FCN opened the gate of segmentation by its inherent advantages. Indeed, FCN has a vital status in Computer-Aided Diagnosis and Treatment of Tumors. Table~\ref{tab3} shows the works using FCN in the clinical practice of tumors.

\begin{table}[htbp]
\centering
\caption{FCN in Computer-Aided Diagnosis and Treatment of Tumors}
\label{tab3}
\setlength{\tabcolsep}{3mm}{
\begin{tabular}{cc}
\toprule
{Tasks} & {Reference}\\
\midrule
{Segmentation} & {\cite{trullo2017segmentation}, \cite{shen2017boundary}, \cite{zhao2018deep}, \cite{christ2017automatic}, \cite{soomro2018automatic}, \cite{drozdzal2018learning}, \cite{bulten2019epithelium}}\\
\bottomrule
\end{tabular}}
\end{table}

FCN was born for semantic segmentation. Hence, it is no surprise that FCN has been a milestone in medical images' segmentation. As Table~\ref{tab3} shows, FCN could segment the CT of liver~\cite{bellver2017detection} and heart~\cite{trullo2017segmentation} (which focus on the dangerous organ segmentation when planning radiotherapy), the MRI of brain~\cite{shen2017boundary, zhao2018deep}, liver~\cite{christ2017automatic}, colorectum~\cite{soomro2018automatic} and prostate~\cite{drozdzal2018learning}.

As a pioneer of the segmentation era, FCN occupies a place in the task of tumor image segmentation and opens a new chapter of the precise definition of tumor contour.

\subsubsection{U-Net}

\begin{figure}[!t]
\begin{center}
      \epsfig{figure=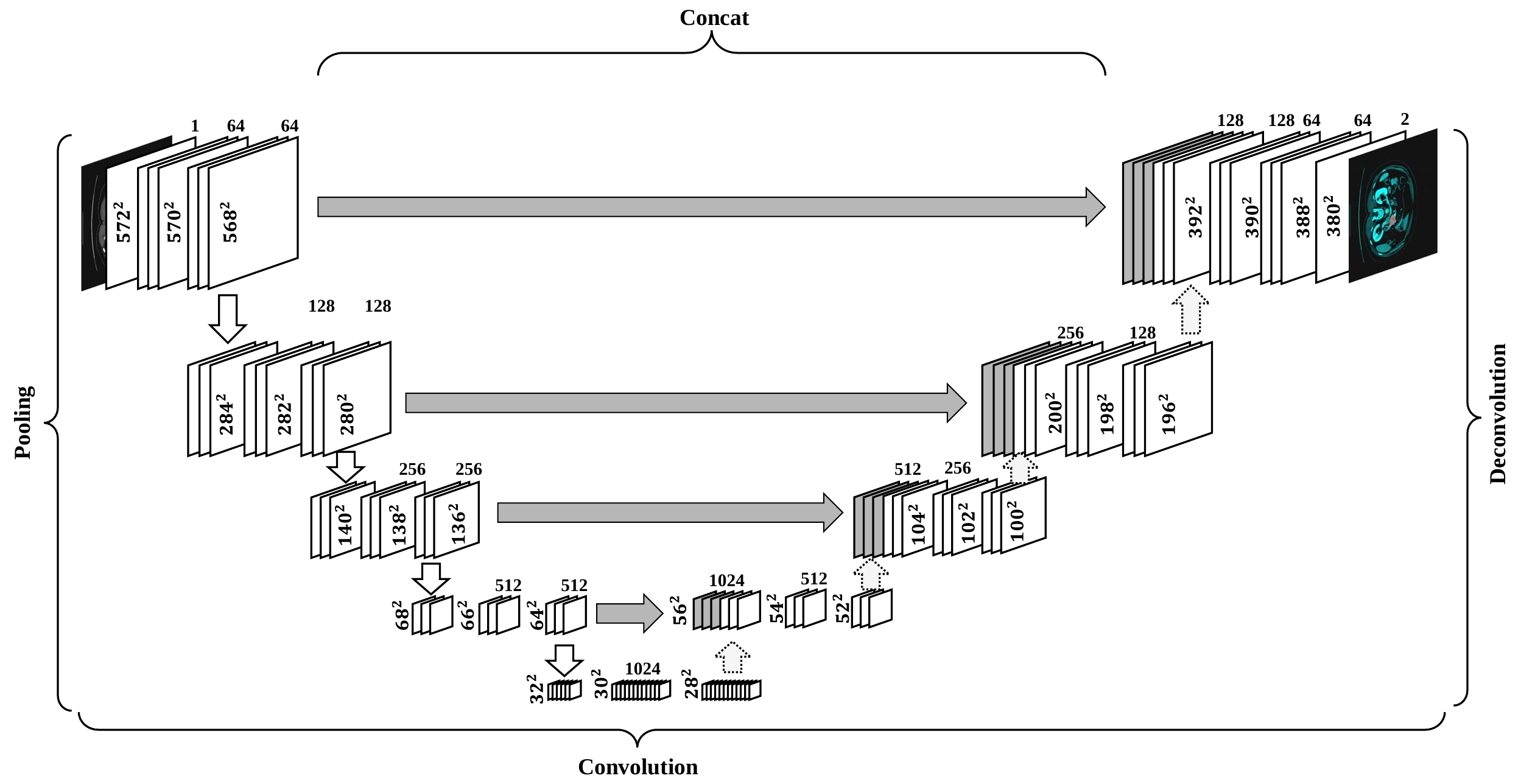,angle=0, width=3.25in}
\end{center}
\caption{The structure of U-Net.}%\vspace{-3mm}
\label{unet}
\end{figure}

As the extension of FCN, U-Net is arguably the advanced deep learning model that improved based on medical data.
U-Net inherits all advantages form FCN but improves the number of deconvolution layer and skip connection layer~\cite{ronneberger2015u}. Also, it has a symmetrical U-shape shown in Figure~\ref{unet}. The left side of U-Net is a contractive path, which extracts features by convolution and max pooling. Also, the right side is an expansive path, which combines the feature map from the left side using concat and sampling the feature map.

U-Net focuses on the characteristics of medical images, which could be summarized as a simple image structure, large image size, a small amount of data, multi-mode, and interpretability. Thus, having the contractive path to capture semantics and the expansive path to locate accurately, U-Net is suitable for the segmentation of the large image, such as histology images (which usually 2GB an image). Moreover, the two paths amount to a trade-off between higher resolution and more abstract features. There are recent works in Table~\ref{tab4} using U-Net in Computer-Aided Diagnosis and Treatment of Tumors.

\begin{table}[htbp]
\centering
\caption{U-Net in Computer-Aided Diagnosis and Treatment of Tumors}
\label{tab4}
\setlength{\tabcolsep}{3mm}{
\begin{tabular}{cc}
\toprule
{Tasks} & {Reference}\\
\midrule
{Segmentation} & {\cite{falk2019u}, \cite{beers2017sequential}, \cite{isensee2017brain}, \cite{guo2018deep}, \cite{zhong20183d}, \cite{bulten2019epithelium}}\\
{Prediction} & {\cite{nguyen2017dose}, \cite{nguyen2018three}, \cite{nguyen2019feasibility}, \cite{nguyen20193d}, \cite{falk2019u}}\\
{Detection} & {\cite{falk2019u}}\\
\bottomrule
\end{tabular}}
\end{table}

The most important uses of U-Net are semantic segmentation, either. However, its excellent performance also could support subsequent tasks. In segmentation, Beers et al.~\cite{beers2017sequential} and Isensee et al.~\cite{isensee2017brain} focuses on brain's MRI, Bulten et al.~\cite{bulten2019epithelium} works on prostate's histology image, Falk et al.~\cite{falk2019u} researches the microscopic video, Guo et al.~\cite{guo2018deep} focus on pancreas's CT , and Zhong et al.~\cite{zhong20183d} uses CT and PET of lung. Furthermore, Falk et al.~\cite{falk2019u} also finished the detection of biomarkers, and Isensee et al.~\cite{isensee2017brain} realized survival prediction in radiobiology.

Moreover, there are other researchers such as Nguyen et al. uses U-Net for prediction. In brief, these works respectively studied prostate~\cite{nguyen2017dose, nguyen2019feasibility} and head and neck (H\&N)~\cite{nguyen2018three, nguyen20193d} on radiotherapy dose prediction based on the data of treatment plan.

Have to say, U-Net is a significant breakthrough in improving the deep learning model based on medical data, which improvement is according to the characteristics of simple medical data structure, small size, and few kinds of targets. This optimization for task objectives also further promotes the exploration of fine segmentation in the field of computer science. So, four years later, u-net is still firmly in the top spot in both medicine and computer science. This phenomenon is very thought-provoking.

\subsection{R-CNN Series}

\begin{figure}[!t]
\begin{center}
      \epsfig{figure=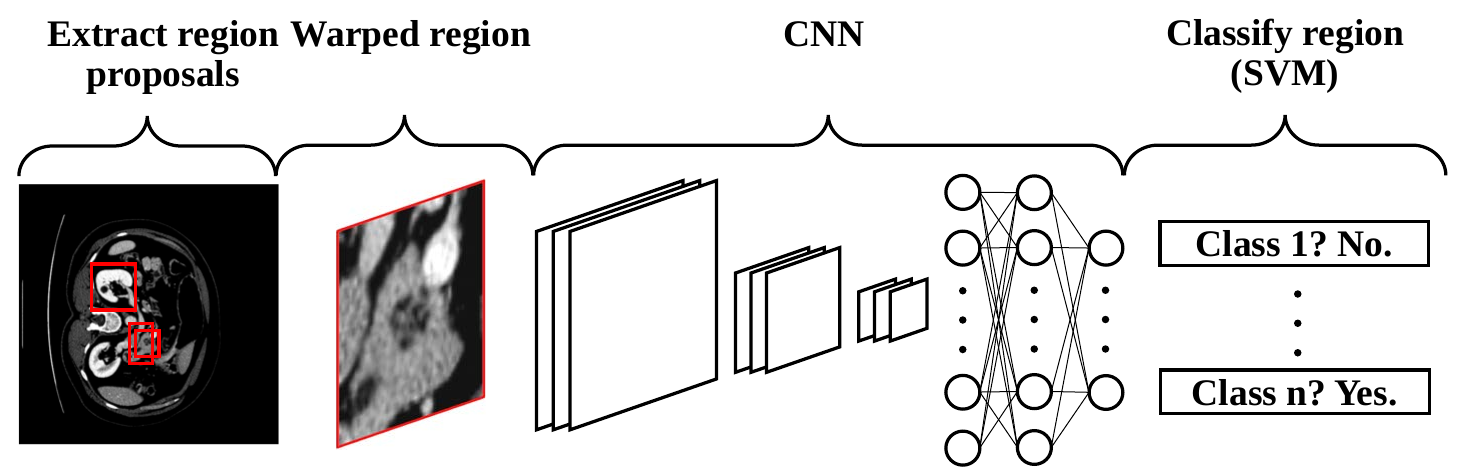,angle=0, width=3.25in}
\end{center}
\caption{The structure of R-CNN.}%\vspace{-3mm}
\label{rcnn}
\end{figure}

Region-CNN (R-CNN) series is a two-stage method that selects proposals on the image first and fixes bounding boxes based on the proposal. As a primitive model, R-CNN~\cite{girshick2014rich} is divided into three parts, finding the candidate box, using CNN to extract feature vectors, and using Support Vector Machine (SVM~\cite{burges1998tutorial}) to classify feature vectors. Figure~\ref{rcnn} shows the specific method, which locates 2000 candidate boxes of objects in the input picture and extracts feature vectors of images in each candidate box, next, classifies and identifies objects in each candidate box.

The improvement of R-CNN is focussing on the training speed. The Fast R-CNN~\cite{girshick2015fast} improves the multi-stage pipeline training of R-CNN and solves the urgent problem for time and space of training. There are two improvements in this work, adding an ROI pooling layer after the last convolution layer, and using a multi-task loss function to add frame regression directly to the training of the CNN network. Moreover, Faster RCNN~\cite{ren2015faster} can view as Fast R-CNN work with the Region Proposal Network (RPN~\cite{ren2015faster}). The corresponding work summarizes in Table~\ref{tab5}.

\begin{table}[htbp]
\centering
\caption{R-CNNs in Computer-Aided Diagnosis and Treatment of Tumors}
\label{tab5}
\setlength{\tabcolsep}{3mm}{
\begin{tabular}{cc}
\toprule
{Tasks} & {Reference}\\
\midrule
{Detection} & {\cite{li2018improved}, \cite{rao2018mitos}, \cite{cai2019efficient}, \cite{akselrod2019cnn}, \cite{ribli2018detecting}}\\
{Classification} & {\cite{akselrod2019cnn}, \cite{ribli2018detecting}}\\
\bottomrule
\end{tabular}}
\end{table}

Detection is the most basic function of the R-CNN series, and it is easy to see that the work based on R-CNN has completed the detection work very well. For instance, Li et al.~\cite{li2018improved} uses Faster R-CNN to detect thyroid ultrasound images. Rao et al.~\cite{rao2018mitos} works on R-CNN and Cai et al.~\cite{cai2019efficient} research Faster R-CNN to detect breast histology images. Also, Faster R-CNN in the researches of Akselrod et al.~\cite{akselrod2019cnn} and Ribli et al.~\cite{ribli2018detecting} are both using for breast mammography to detect and classify malignant or benign lesions.

As the pioneer of target detection, R-CNNs, the two-stage method, has completed the detection of biomarkers well in the process of tumor diagnosis and treatment. It is worth mention that in the field of computer science, there are actually other object detection methods that are being updated, such as You Only Look Once series (YOLO~\cite{redmon2016you}, YOLOv2~\cite{redmon2017yolo9000}, YOLOv3~\cite{redmon2018yolov3}), and \textit{Segmentation is all you need}~\cite{wu2019segmentation} which written by Wu et al. in 2019. The reason why the above works are not described in detail here is that the YOLO series, as one-stage methods, has lost parts of precision (which is a crucial indicator in tumor diagnosis and treatment). Meanwhile, the work named \textit{Segmentation is all you need}, which is the latest and most breakthrough deep learning method for object detection that has not been the star in the medical field. However, one of the advantages of the work finished by Cheng et al. is the detection of difficult small targets, and its performance in subsequent medical tasks is expected.

\section{Recent Work in Intelligent Diagnosis and Treatment of Tumors}
\label{sec:Recentwork}

\subsection{In-Vitro Diagnosis (IVD)}

In-Vitro Diagnosis (IVD) is the first stage of tumor diagnosis, which is mainly responsible for the detection and screening of tumor markers or tumor characteristics. Early detection of cancer often determines the prognosis of patients' quality of life or even life. It is the crucial reason why deep learning is needed at this stage. IVD uses three kinds of data, which will introduce as follow.
\begin{enumerate}
\item Images or videos of cells in blood and tissue fluid generated from microscope to discover if there exists Circulating Tumor Cell (CTC).
\item Indicators in biophysics and biochemistry from Flow Cytometer (FCM) to discover the specified subset of cells.
\item Gene expression by arrays obtained from the public datasets to discover if there exists a gene disorder.
\end{enumerate}
Currently, there are public datasets online with labels in this stage. Such as, National Cancer Institute Genomic Data Commons (GDC) Data Portal (see: \textbf{\url{https://portal.gdc.cancer.gov/}}), Genomics of Drug Sensitivity in Cancer (GDSC, see: \textbf{\url{https://www.cancerrxgene.org/}}), MICCAI Challenge on Circuit Reconstruction from Electron Microscopy Images (CREMI, see: \textbf{\url{https://cremi.org/data/}}), and LYmphocyte aSsessmenT hackathOn (LYSTO, see: \textbf{\url{https://lysto.grand-challenge.org/}}). The vast majority of open datasets can be found at Grand Challenge (see: \textbf{\url{https://grand-challenge.org/challenges/}}), or even the data types to be introduced in the next three stages in Computer-Aided Diagnosis and Treatment of Tumors.

%such as International Symposium on Biomedical Imaging (ISBI), MICCAI Challenge on Circuit Reconstruction from Electron %Microscopy Images (CREMI), and LYmphocyte aSsessmenT hackathOn (LYSTO).

The main clinical problems at this stage are the early screening of tumors, identifying tumor stages and subtypes, monitoring the efficacy of therapy, and predicting prognosis of tumors. All of these medical tasks can be seen as a classification task in computer science. Of course, segmentation and object detection can also accomplish the above tasks if it is necessary. Now, with the development of CNN, there is a practical approach to solve the above problems in deep learning. Table~\ref{IVD} summarizes recent work for CNN in IVD.

\begin{table}[ht]
  \caption{Recent work in In-Vitro Diagnosis}
  \centering
  \label{IVD}
  \begin{tabular}{c|c|c}
  \hline
  {Data} & {Tasks}  & {Reference} \\
  \hline
  \multirow{2}{*}{Image} & {Detection} & {\cite{falk2019u}}\\
  \cline{2-3}
   & {Segmentation} & {\cite{falk2019u}, \cite{tran2018blood}, \cite{drozdzal2018learning}}\\
  \hline
  {Indicators} & {Classification} & {\cite{doan2018diagnostic}}\\
  \hline
  {Gene} & {Prediction} & {\cite{chang2018cancer}, \cite{xiao2018deep}}\\
  \hline
  \end{tabular}
\end{table}

\subsubsection{Image data diagnosis in IVD}

\begin{figure}[!t]
\begin{center}
      \epsfig{figure=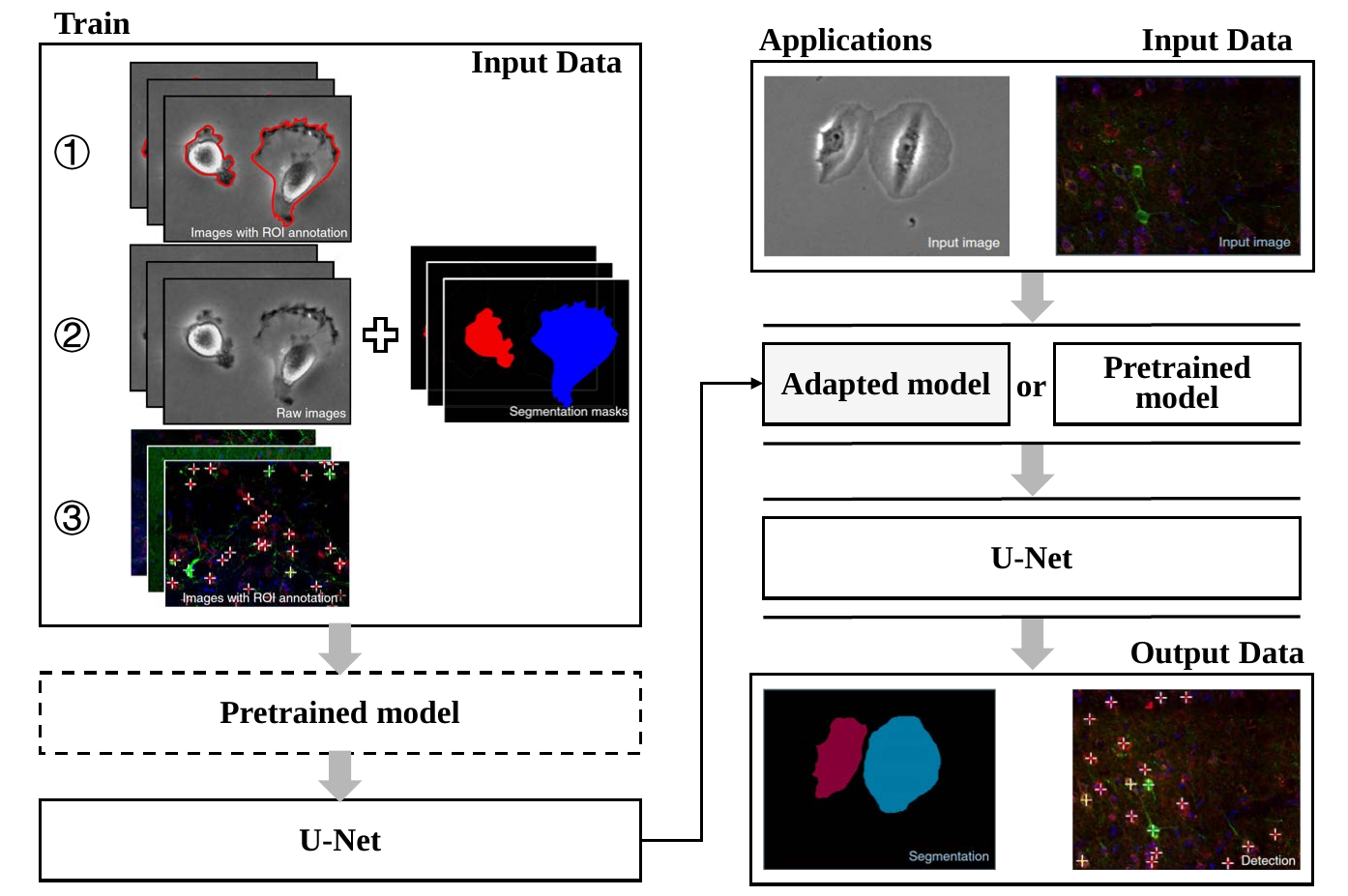,angle=0, width=3.25in}
\end{center}
\caption{The pipeline of the research worked by falk et al.~\cite{falk2019u} which is basically different annotation types of data learning using U-Net.}%\vspace{-3mm}
\label{I1}
\end{figure}

In detection, the work of Falk et al.~\cite{falk2019u} is excellent by using U-Net, which enables non-machine-learning experts to analyze their data and could save manual annotation effort in a wide variety of quantification tasks. Also, this work supports single-cell segmentation with conditional adaptability. Furthermore, the datasets in this work (such as F1-MSC, F2-GOWT1, F3-SIM, F4-HeLa, DIC1-HeLa, PC1-U373, and PC2-PSC) are from the International Symposium on Biomedical Imaging (ISBI) Cell Tracking Challenge 2015. Figure~\ref{I1} shows the pipeline of this work. Firstly, training the U-net on the local machine, a dedicated remote server, or a cloud service. After that, the adaptation of U-Net to newly annotated data by using transfer learning. This work performance well in follows tests.
\begin{itemize}
\item Detection (2D) of colocalization in two-channel epifluorescence images.
\item Detection (3D) of fluorescent-protein-tagged microglial cells in a five-channel confocal-image.
\item Segmentation (2D) of morphometric cell description from fluorescence, differential interference contrast, phase-contrast, and bright-field microscopy.
\item Segmentation (3D) of 3D bright-field images and neurite is tracing in electron microscopy images.
\end{itemize}

\begin{figure}[!t]
\begin{center}
      \epsfig{figure=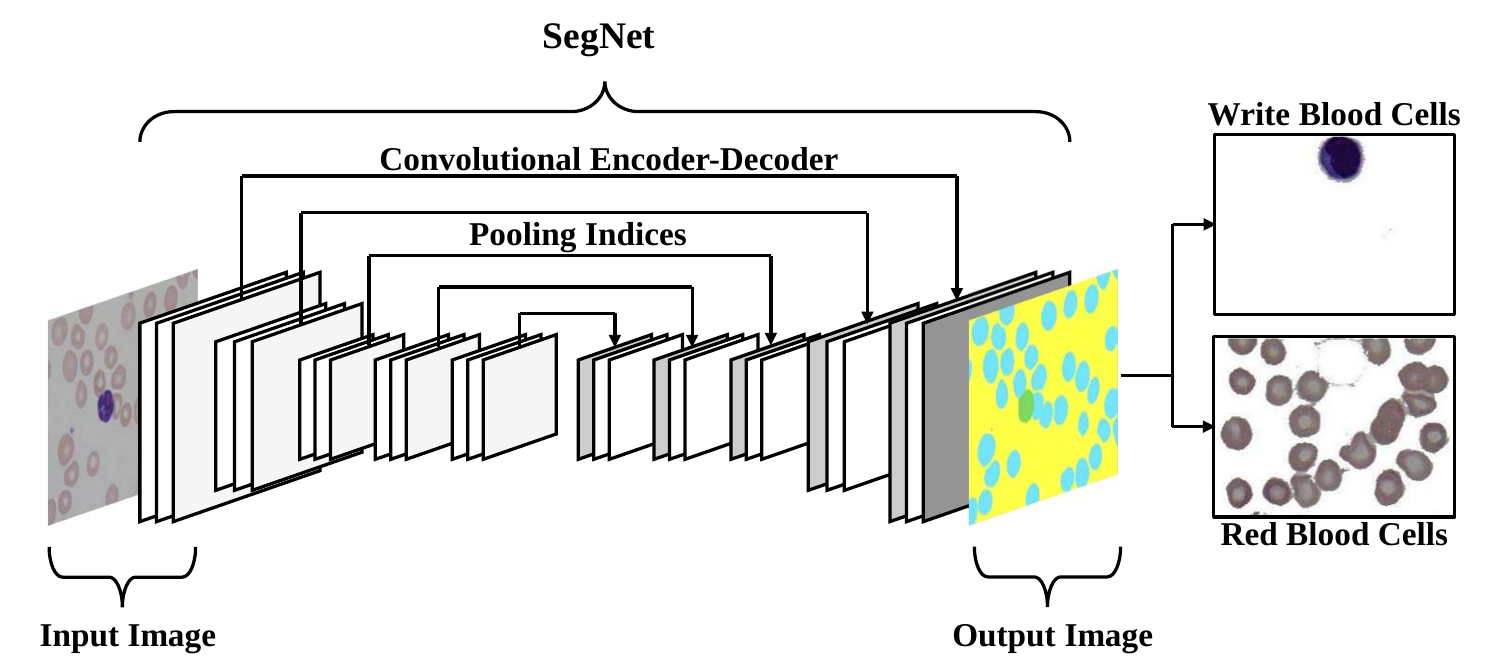,angle=0, width=3.25in}
\end{center}
\caption{The pipeline of~\cite{tran2018blood} depends on SegNet to segment WBCs and RBCs.}%\vspace{-3mm}
\label{I12}
\end{figure}

In segmentation, Tran et al.~\cite{tran2018blood} uses SegNet (which is a variant of FCN~\cite{badrinarayanan2017segnet}) to efficiently segment both Red Blood Cells (WBCs) and White Blood Cells (RBCs). The dataset is the peripheral blood smear images from the ALL-IDB1 database. The details of this work are shown in Figure~\ref{I12}. The first step utilizes SegNet to label all of WBCs and RBCs in input images with different colors. Then, it separated all leucocytes and erythrocytes into two individual images. Finally, using the result images for various purposes, such as early detection of leukemia (a type of white blood cell cancer) and count of the complete blood cell.

\subsubsection{Gene data diagnosis in IVD}

Now, genomic data is generally used for prediction, such as cancer prediction and anticancer drug responsiveness prediction.

\begin{figure}[!t]
\begin{center}
      \epsfig{figure=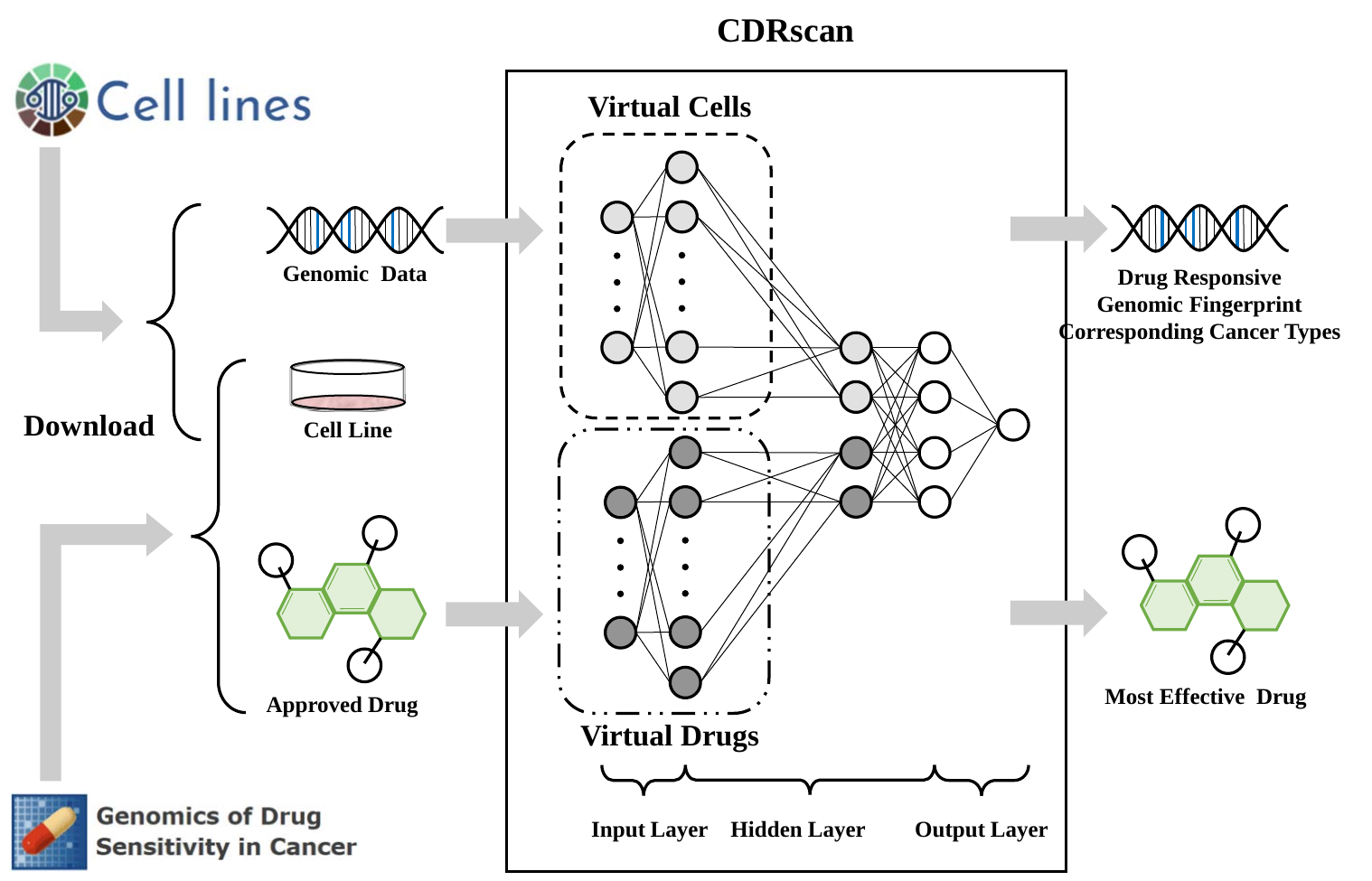,angle=0, width=3.25in}
\end{center}
\caption{The pipeline of the work written by Chang et al.~\cite{chang2018cancer} depends on two-step CNN to predict anticancer drugs responsiveness.}%\vspace{-3mm}
\label{I13}
\end{figure}

There is a standard work written by Chang et al.~\cite{chang2018cancer} in the Table~\ref{IVD} researches for drug efficacy prediction, which employs CNN to process the genomic mutational fingerprints of cell lines and the molecular fingerprints of drugs (these data comes from CCLP1, GDSC6, CGC, GDSC). It may allow the selection of the most effective anticancer drugs for the genomic profile of the individual patient in the future. The details will show in Figure~\ref{I13}. In this research, the model consists of two parts, a dual convergence CNN and a generalizable prediction model. Furthermore, the inputs of the model can be molecular information of a particular small molecule, and the model could predict which of Genomics in Drug Sensitivity in Cancer (GDSC) anticancer drugs would be valid.

\subsubsection{Indicators data diagnosis in IVD}

\begin{figure}[!t]
\begin{center}
      \epsfig{figure=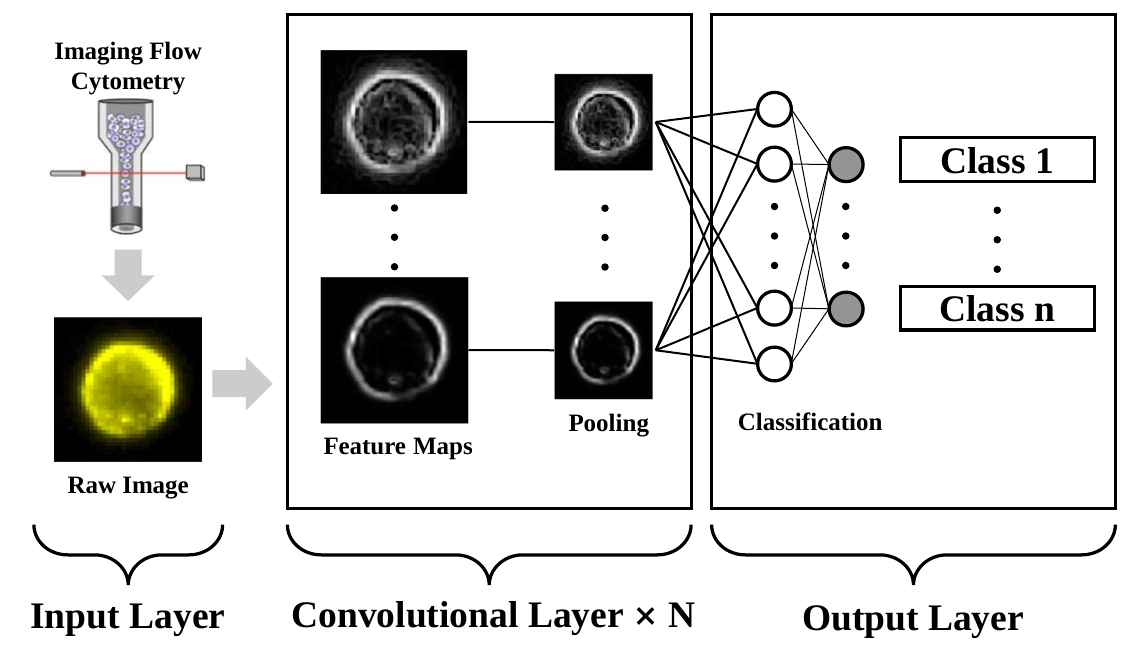,angle=0, width=3.25in}
\end{center}
\caption{The illustration of the work written by Doan et al.~\cite{doan2018diagnostic} depends on CNN to analysis the cells by classification. For instance, Class1 and 2 could correspond to leukemic and normal cells.}%\vspace{-3mm}
\label{I14}
\end{figure}

With the increasing maturity of deep learning methods, more and more types of data are expected to be processed by deep learning methods, and Doan et al.~\cite{doan2018diagnostic} shown in Table~\ref{IVD} is one of them. Doan et al. analysis the diagnostic potential of Imaging Flow Cytometry (FCM), which database contains the image of each cell and the quantified multiple properties constituents of interest indicators (including proteins, nucleic acids, glycolipids) in multiple subcellular compartments (nucleus, mitochondria). Figure~\ref{I14} is the illustration of the assumption by using general CNN in the research. This work may use to analyze the rare cell types such as circulating tumor cells (which are cancer cells that escaped from a primary tumor and circulate in the bloodstream) and transition states, such as cell cycle phases (mitosis).

As a summary, even the In-Vitro Diagnosis is the earliest stage for the diagnosis of tumors, the data types and the large data sources are pool. There are several reasons which based on the clinical practice for In-Vitro Diagnosis mainly causes it.
\begin{itemize}
\item Many cytology reports do not require the same photographic evidence as imaging diagnosis, so doctors have no needs to gather image data during the diagnosis; instead, they record the values detected.
\item Currently, general health examination conducted by hospitals do not include the genomic testing of tumors, and so do the medical institutions. Also, most patients lack awareness of genetic risk prediction. Therefore, to obtain a large number of high-quality tumor genome screening data is difficult.
\item Clinical indicator data are presented in the form of text, and Natural Language Processing (NLP) is required to complete the task of text abstract before further analysis. However, text abstract is rarely used in medical reports, which results in a lack of available data.
\end{itemize}

\subsection{Imaging Diagnosis (ID)}

If abnormal discoveries are finding in IVD or simply because of the physical discomfort reaching the diagnostic range, patients may have to follow a doctor's orders to finish imaging diagnosis.
The Imaging Diagnosis uses medical images to discover if there exist leisures and dangerous organs. These images come from  UB, X-ray, CT, MRI, Digital Subtraction Angiography (DSA), RadioNuclide Imaging (RNI), Positron Emission Computed Tomography (PET), and Single-Photon Emission Computed Tomography (SPECT).
There are plenty of public datasets about medical images. As usual, these datasets could be found in Grand Challenge, such as Kidney Tumor Segmentation Challenge (KiTS), Combined Healthy Abdominal Organ Segmentation (CHAOS), Decathlon-10. Furthermore, The Cancer Imaging Archive (TCIA, see: \textbf{\url{https://www.cancerimagingarchive.net/}}) contains packaged datasets, such as multi-modal or multi-type data of tumors. Of course, some datasets have histology images, multiple types of medical images, genomic data, and clinical information, which are suitable for the integration of multi-modal data and tumor diagnosis and treatment.

In this stage, the essential clinical demands include diagnosing, detecting, and delineating the tumors or dangerous organs. Thanks to the applications supported by CNN, FCN, U-net, and R-CNNs, most of the problems in Imaging Diagnosis could be solved in deep learning. Among these models, CNN could be used to diagnose tumors and extract features before delineating the tumors; FCN and U-Net mostly be used to delineate the tumors; R-CNNs are good at object detection for tumors. Recent works are shown in Table~\ref{ID}.

\begin{table}[ht]
  \caption{Recent work in Imaging Diagnosis}
  \centering
  \label{ID}
  \setlength{\tabcolsep}{3mm}{
  \begin{tabular}{c|c|c}
  \hline
  {\bf Data} & {\bf Tasks} & {\bf Reference}\\
  \hline
  \multirow{2}{*}{MRI} & Segmentation & \tabincell{c}{\cite{saouli2018fully}, \cite{perkuhn2018clinical}, \cite{kamnitsas2017ensembles}, \cite{milletari2017hough},\\ \cite{laukamp2019fully}, \cite{trebeschi2017deep}, \cite{li2018tumor}, \cite{kao2018brain},\\ \cite{shen2017boundary}, \cite{zhao2018deep}, \cite{christ2017automatic}, \cite{soomro2018automatic},\\ \cite{drozdzal2018learning}, \cite{isensee2017brain}, \cite{beers2017sequential}, \cite{li2017deep}} \\
  \cline{2-3}
   & {Prediction} & {\cite{li2017deep}, \cite{kao2018brain}, \cite{isensee2017brain}}\\
  \hline
  \multirow{2}{*}{CT} & {Segmentation} & \tabincell{c}{\cite{bellver2017detection}, \cite{trullo2017segmentation}, \cite{christ2017automatic}, \cite{drozdzal2018learning},\\ \cite{guo2018deep}, \cite{ardila2019end}}\\
  \cline{2-3}
   & {Prediction} & \tabincell{c}{\cite{jackson2018deep}, \cite{foote2018real}, \cite{oakden2017precision}, \cite{wang2019deep},\\ \cite{zhao20183d}\, \cite{ardila2019end}}\\
  \hline
  \multirow{2}{*}{PET} & {Segmentation} & {\cite{leung2018deep}, \cite{zhong20183d}} \\
  \cline{2-3}
   & {Prediction} & {\cite{zhang2017personalized}}\\
  \hline
  \multirow{2}{*}{Mammography} & {Detection} & \multirow{2}{*}{\cite{akselrod2016region}, \cite{ribli2018detecting}}\\
  \cline{2-2}
   & {Classification} & \\
  \hline
  {US} & {Detection} & {\cite{li2018improved}}\\
  \hline
 \end{tabular}}
\end{table}

\subsubsection{MRI diagnosis in ID}

MRI has good discrimination of soft tissue and no ionizing radiation damage to the human body, so it is good at imaging tumors in the brain, bladder, rectum, reproductive system, and other parts.

\begin{figure}[!t]
\begin{center}
      \epsfig{figure=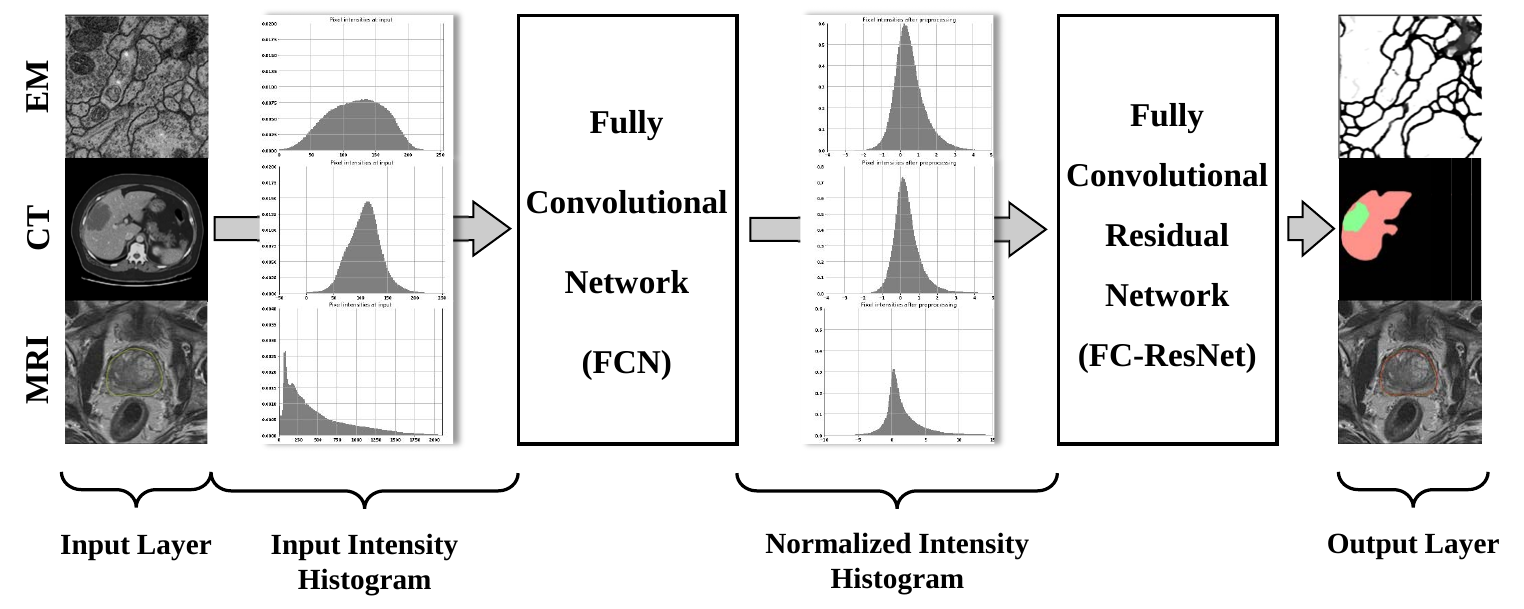,angle=0, width=3.25in}
\end{center}
\caption{The illustration of the work written by Drozdzal et al.~\cite{drozdzal2018learning} depends on FCN and FC-ResNet to segment both 2D and 3D medical images, such as EM, CT, and MRI.}%\vspace{-3mm}
\label{I21}
\end{figure}

In segmentation, we have to introduce the work written by Drozdzal et al.~\cite{drozdzal2018learning}, which combines Fully Convolutional Networks (FCNs) with Fully Convolutional Residual Networks (FC-ResNets~\cite{drozdzal2016importance}) to segment medical images, such as electron microscopy (EM) image, CT of the liver, and MRI of the prostate. The results reveal that this model is working well in both 2D and 3D medical images, which means that it could achieve accurate segmentations on a variety of image modalities and different anatomical regions. Details are shown in Figure~\ref{I21}, which uses the FCN to obtain pre-normalized images, and then iteratively refined employing the FC-ResNet to generate a segmentation prediction.

\begin{figure}[!t]
\begin{center}
      \epsfig{figure=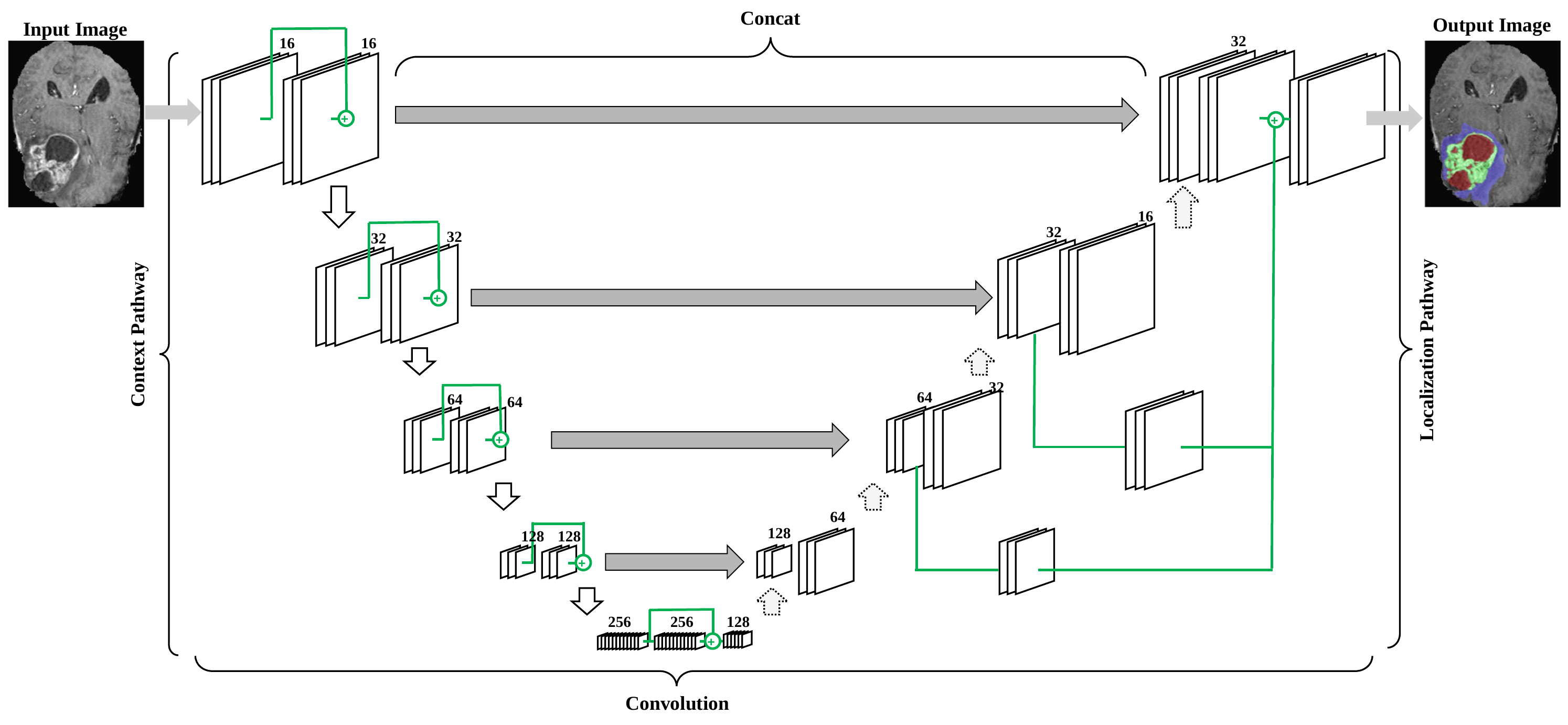,angle=0, width=3.25in}
\end{center}
\caption{The illustration of the work written by Isensee et al.~\cite{isensee2017brain} depends on U-Net to segment brain MRI and RF to predict radiomics survival.}%\vspace{-3mm}
\label{I22}
\end{figure}

In prediction, Isensee et al.~\cite{isensee2017brain} finished the survival prediction by training an ensemble of Random Forest (RF~\cite{breiman2001random}) regressor and multi-layer perceptrons on shape features describing the tumor subregions. Figure~\ref{I22} is the pipeline of the work, which is derived from the U-Net. In short, the context pathway (left) aggregates high-level information that is subsequently localized precisely in the localization pathway (right). Also, this work injects gradient signals deep into the network through depth supervision.

Different from the work mentioned above~\cite{isensee2017brain}, these works~\cite{li2017deep, kao2018brain} also have a prediction part using Support Vector Machine (SVM~\cite{burges1998tutorial}). Noticed that both Random Forest (RF) and Support Vector Machine (SVM) are the machine learning method, maybe there will be a more deep learning method using for prediction soon in this field.

\subsubsection{CT diagnosis in ID}

CT is of high diagnostic value for tumors of the central nervous system, head and neck, chest and abdomen, and pelvic cavity. It is widely used in the diagnosis and treatment of computer-assisted tumors.

\begin{figure}[!t]
\begin{center}
      \epsfig{figure=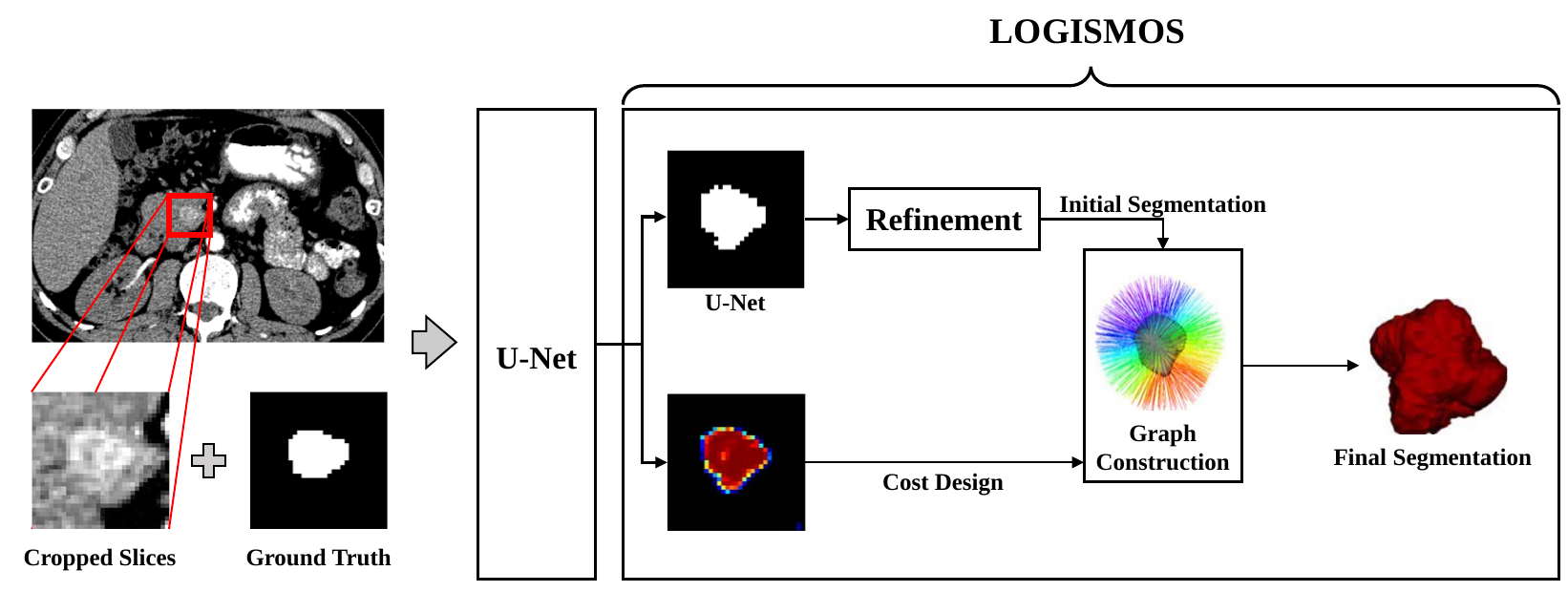,angle=0, width=3.25in}
\end{center}
\caption{The pipeline of the work written by Guo et al.~\cite{guo2018deep} which adopts the U-Net to integrate intra-slice and adjacent-slice contexts, and regulates the 3D shape by LOGISMOS in pancreas tumor segmentation.}%\vspace{-3mm}
\label{I23}
\end{figure}

In segmentation, Guo et al.~\cite{guo2018deep} employs U-Net which refined by Gaussian Mixture Model (GMM) and morphological operations to segment 3D pancreas tumor. To finish the morphological operations, the authors adopt a model named LOGISMOS, which is a graph-based framework that translates geometric constraints of interacting surfaces and objects into graph arcs and the likelihood of segmentation surface positioning into graph node/arc costs. The details are shown in Figure~\ref{I23}, which combines U-Net (to integrate intra-slice and adjacent-slice contexts) and LOGISMOS (to regulate the 3D shape) for 3D tumor segmentation.

\begin{figure}[!t]
\begin{center}
      \epsfig{figure=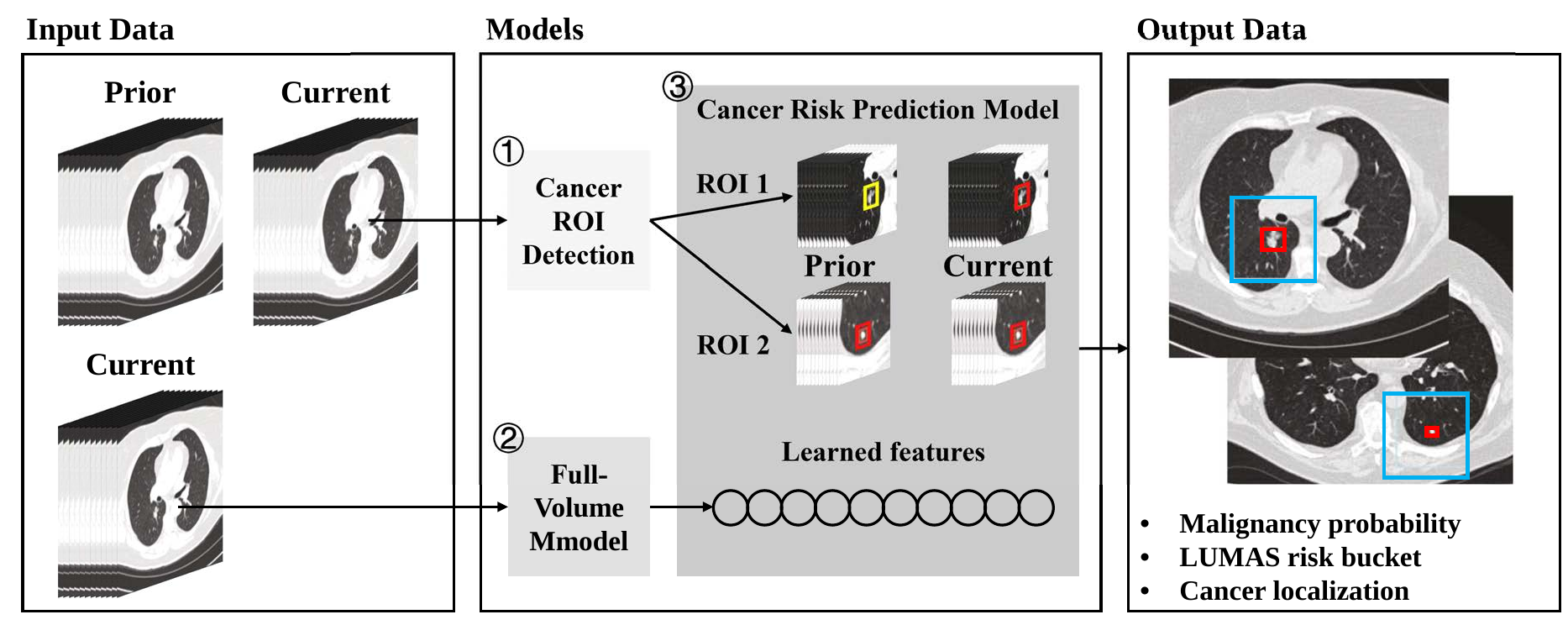,angle=0, width=3.25in}
\end{center}
\caption{The pipeline of the work of Ardila et al.~\cite{ardila2019end} which is basically based on CNN.}%\vspace{-3mm}
\label{I2}
\end{figure}

In prediction, Ardila et al.~\cite{ardila2019end} built an end-to-end approach based on CNN for CT images, which outputs overall malignancy prediction for the case, a risk bucket score (LUMAS) and localization for predicted cancerous nodules. The pipeline shows in Figure~\ref{I2}, which contains three parts based on CNN. There is a full-volume model (to perform end-to-end analysis of LDCT), cancer Region of Interest (ROI) detection model (to detect 3D cancer candidate regions), and cancer risk prediction model (to operate on outputs from above two models). This work achieved an outstanding result, about 94.4\% area under the curve on National Lung Cancer Screening Trial cases (NLST).

\subsubsection{PET diagnosis in ID}

Positron Emission Tomography and Computed Tomography (PET-CT) dual-modality imaging provide critical diagnostic information in current cancer diagnosis and therapy.

\begin{figure}[!t]
\begin{center}
      \epsfig{figure=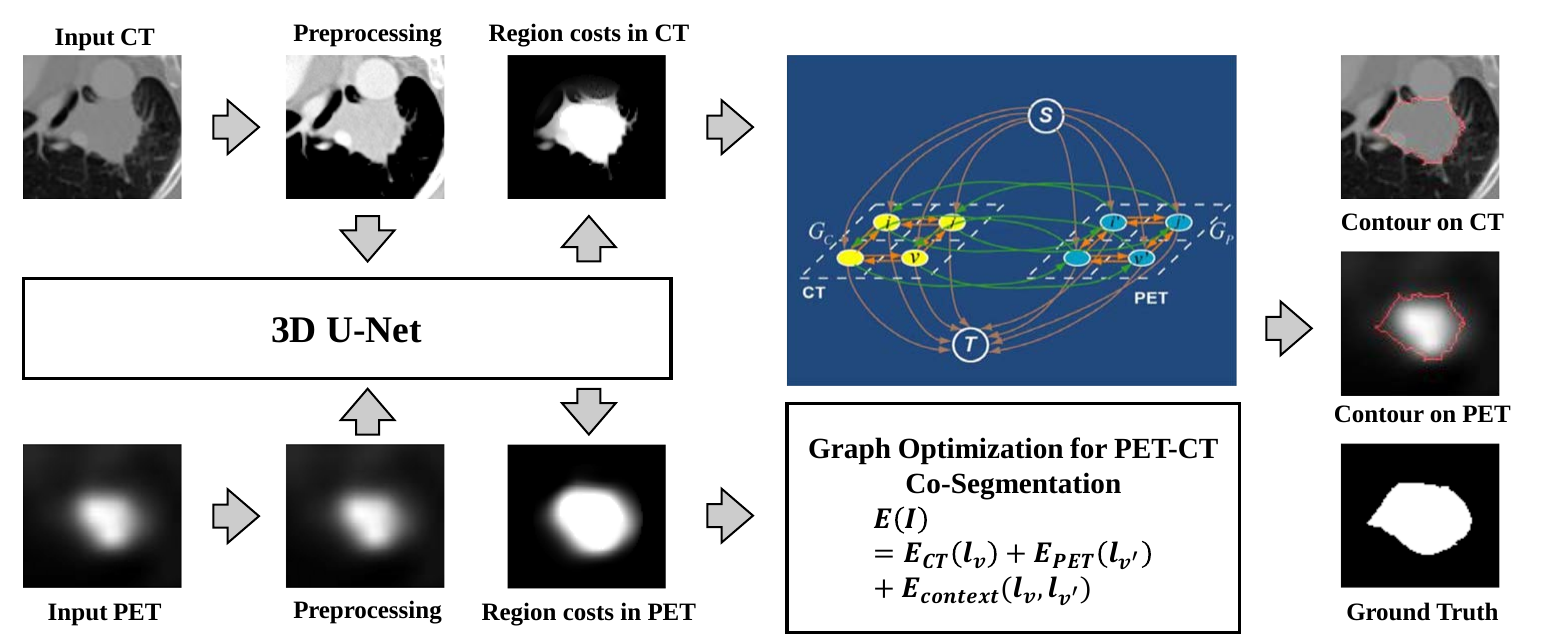,angle=0, width=3.25in}
\end{center}
\caption{The pipeline of the work from Zhong et al.~\cite{zhong20183d} which designed two independent 3D U-Net for PET and CT to produce high-quality regions costs for subsequent graph-based co-segmentation in lung's dataset.}%\vspace{-3mm}
\label{I24}
\end{figure}

In segmentation, Zhong et al.~\cite{zhong20183d} employ a model that combines 3D U-Net, and the graph cut based co-segmentation model to automated, accurate tumor delineation based on PET-CT, which is essential in computer-assisted tumor reading and interpretation. The details are shown in Figure~\ref{I24}. In this work, the researchers focus on lung cancer PET-CT and generate high-quality voxel-level tumor confidences that were further used to locate the tumor boundary with the powerful co-segmentation model.

\begin{figure}[!t]
\begin{center}
      \epsfig{figure=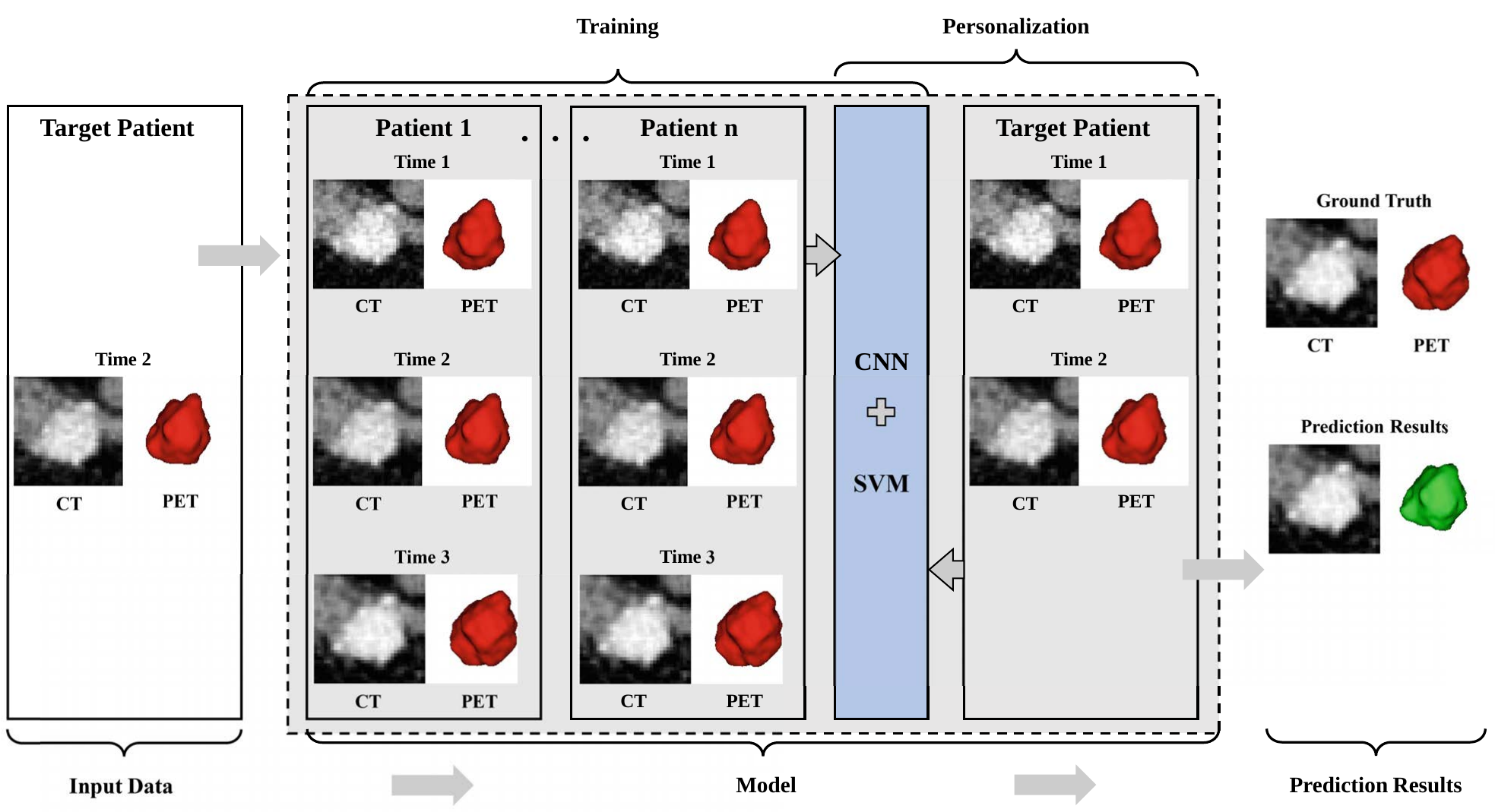,angle=0, width=3.25in}
\end{center}
\caption{The structure of the work from Zhang et al.~\cite{zhang2017personalized} which is a framework of the voxel-wise prediction of tumor growth via CNN and SVM.}%\vspace{-3mm}
\label{I25}
\end{figure}

In prediction, Zhang et al.~\cite{zhang2017personalized} propose a CNN to predict the pancreas tumor growth pattern that incorporates both the population trend and personalized data. Figure~\ref{I25} shows the structure. The deep features extracted by CNN are combined with the time intervals and the clinical factors to feed a process of feature selection and after that, selected a robust feature subset by the Support Vector Machine Recursive Feature Elimination (SVM RFE). Finally, a SVM predictive model was used to predict the tumor's spatiotemporal growth and progression.

\subsubsection{Mammography diagnosis in ID}

\begin{figure}[!t]
\begin{center}
      \epsfig{figure=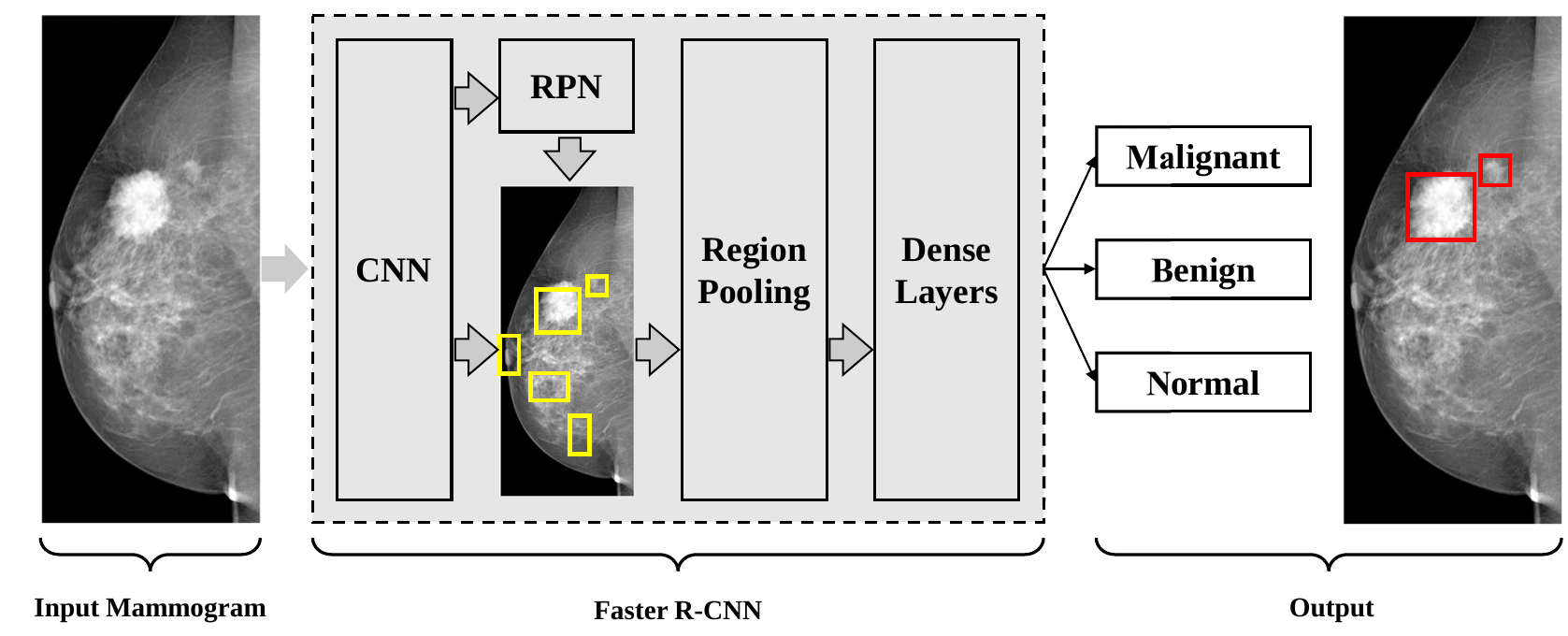,angle=0, width=3.25in}
\end{center}
\caption{The structure of the work from Ribli et al.~\cite{ribli2018detecting} which is a framework of the object detection in tumor early screening via Faster R-CNN and SVM.}%\vspace{-3mm}
\label{I26}
\end{figure}

Now, mammography only uses for breast screening. Thus most of the researchers focus on the detection and classification of early cancer. As Figure~\ref{I26} shown, Ribli et al.~\cite{ribli2018detecting} adopt Faster R-CNN by optimized both the object detection and classifier part to detect and classify malignant or benign lesions on mammogram without any human intervention. Also, the mammogram images come from the public INbreast database. This research could realize computer-aided detection in mammographic screening.

\subsubsection{Ultrasound (US) diagnosis in ID}

Ultrasound (US) is the most widely used tumor screening method, thanks to its non-radioactive, multi-directional cross-sectional imaging, real-time dynamic, easy to operate, fast, and other characteristics. However, ultrasound needs the assistance of deep learning because of its disadvantages of lack of specificity, focus on local areas, and is easily influenced by doctors' experience.

\begin{figure}[!t]
\begin{center}
      \epsfig{figure=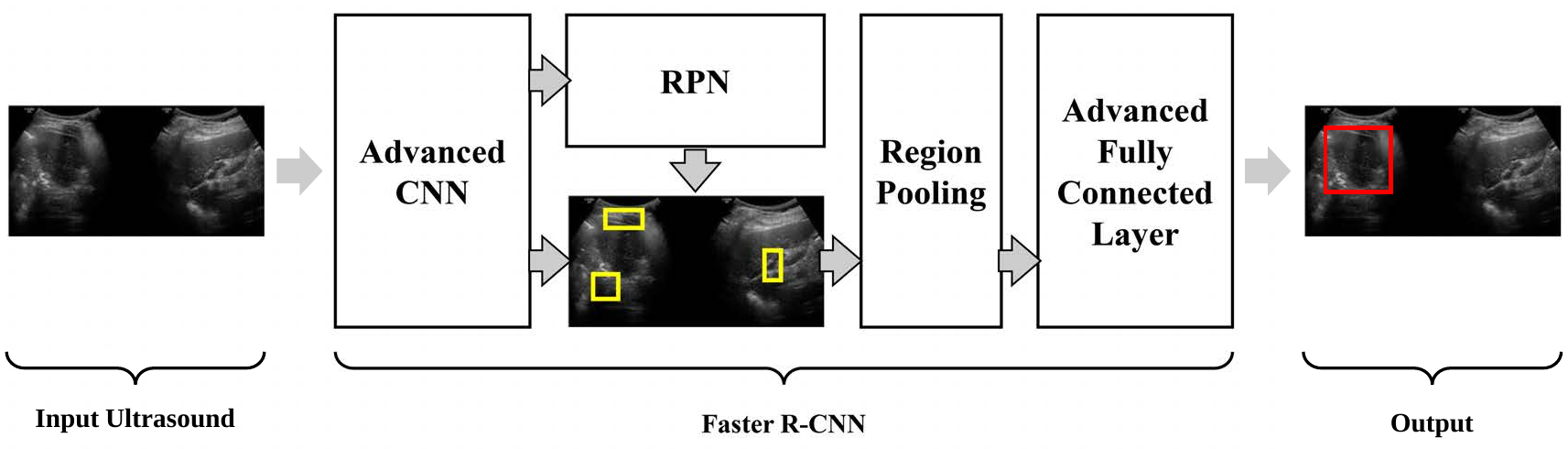,angle=0, width=3.25in}
\end{center}
\caption{The structure of the work from Li et al.~\cite{li2018improved} which adopt Faster R-CNN to detect thyroid papillary carcinoma.}%\vspace{-3mm}
\label{I27}
\end{figure}

For instance, Li et al.~\cite{li2018improved} employs Faster R-CNN to add a spatial constrained layer before the output layer of CNN, as the pipeline is shown in Figure~\ref{I27}, which concatenated and normalized the conv3 and conv5 layer and then add a spatial constrained layer before the output layer. This work is more suitable for thyroid papillary carcinoma detection in ultrasound images, and the property of classification is better than SVM. This work could improve the ability of screening cancer in thyroid papillary carcinoma images fast.

As a summary, Image Diagnosis runs through the whole process of tumor diagnosis and treatment. Its data is easy to obtain and the data is huge. It is an essential part of the current research on computer-assisted tumor diagnosis and treatment thanks to the rapid development of computer vision. As can be seen from this section, the research on individual organs (especially large organs or organs that are easy to inspect) has reached a saturation point, and the future trend is nothing more than improvement in accuracy and speed. However, targeted studies are still needed for small organ tumors, such as pancreas and ovary. These small organ tumors are high malignancy and a short course of the disease (which means the data is scarcity). Besides, there is still little research on multi-modal data (although such multi-modal databases like TCIA has existed).

\subsection{Pathological Diagnosis}

Pathological Diagnosis is the gold standard in the tumor's diagnosis and treatment, which determines the stage and subtype of tumors. Moreover, treatment planning depends on the results of the Pathological Diagnosis. However, because of its invasive nature, Pathological Diagnosis can not be used for early screening.

Pathological Diagnosis based on deep learning uses the images of leisures, which prepared by histopathological methods (such as H\&E-stained) to discover the exceptions on the tissue structure, cell morphology, and growth pattern of the lesion. Now, the amount of public datasets is available. As usual, Grand Challenge and TCIA are available.

This stage has to screen the cell nucleus of tumors, delineate the pathological leisure, and detect the object of tumors. Same as Imaging Diagnosis, which input data is the image, and deep learning methods (such as CNN, FCN, U-Net, and R-CNNs) work well on Pathological Diagnosis. Follow Table~\ref{PD} is the collection of recent works.

\begin{table}[ht]
  \caption{Recent work in Pathological Diagnosis}
  \centering
  \label{PD}
  \setlength{\tabcolsep}{3mm}{
  \begin{tabular}{c|c|c}
  \hline
  {Data} & {Tasks}  & {Reference} \\
  \hline
  \multirow{3}{*}{\tabincell{c}{Histology\\images}} & {Classification} & \tabincell{c}{\cite{araujo2017classification}, \cite{cruz2017accurate}, \cite{saltz2018spatial}, \cite{golatkar2018classification},\\\cite{khoshdeli2018deep},\cite{klein2019maldi}, \cite{khosravi2018deep}, \cite{zhao20183d}}\\
  \cline{2-3}
   & {Segmentation} & {\cite{bulten2019epithelium}, \cite{janowczyk2018resolution}, \cite{coudray2018classification}, \cite{zhao20183d}} \\
  \cline{2-3}
   & {Prediction} & {\cite{schaumberg2018h}, \cite{bychkov2018deep}, \cite{zhao20183d}} \\
  \hline
  \end{tabular}}
\end{table}

\begin{figure}[!t]
\begin{center}
      \epsfig{figure=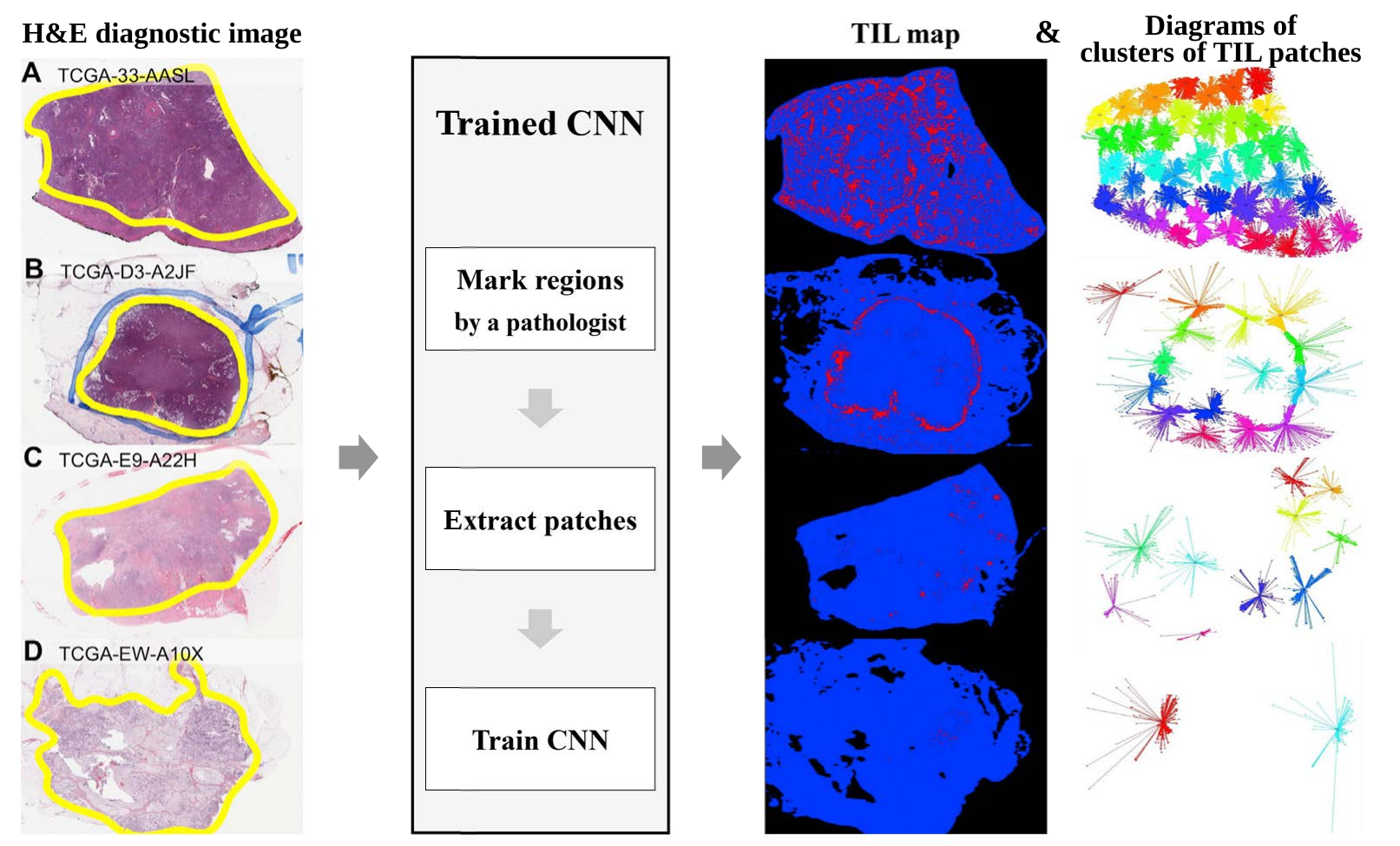,angle=0, width=3.25in}
\end{center}
\caption{The pipeline of the work from Saltz et al.~\cite{saltz2018spatial} which could realize a global structure classification in lymphocytes by CNN.}%\vspace{-3mm}
\label{P2}
\end{figure}

In Classification, Saltz et al.~\cite{saltz2018spatial} is standard research that using CNN to classify 13 cancer types in lymphocytes on pathology images compared with The Cancer Genome Atlas (TCGA). In this research, as Figure~\ref{P2} shows, each patch was annotated with a necrosis region mask segmented by a pathologist, and the expert reviewed and corrected predicted Tumor-Infiltrating lymphocytes (TIL) during the CNN training stage. This effort assess lymphocytic infiltrates across multiple TCGA tumor types for correlation. Genomic and epigenomic assessments of lymphocytic infiltrate, as well as clinical outcome, which utilizes well of TCGA and shows that molecular assessments of TILs (generated by the molecular platforms of the TCGA) can correlate with clinical outcome for specific tumor types.

\begin{figure}[!t]
\begin{center}
      \epsfig{figure=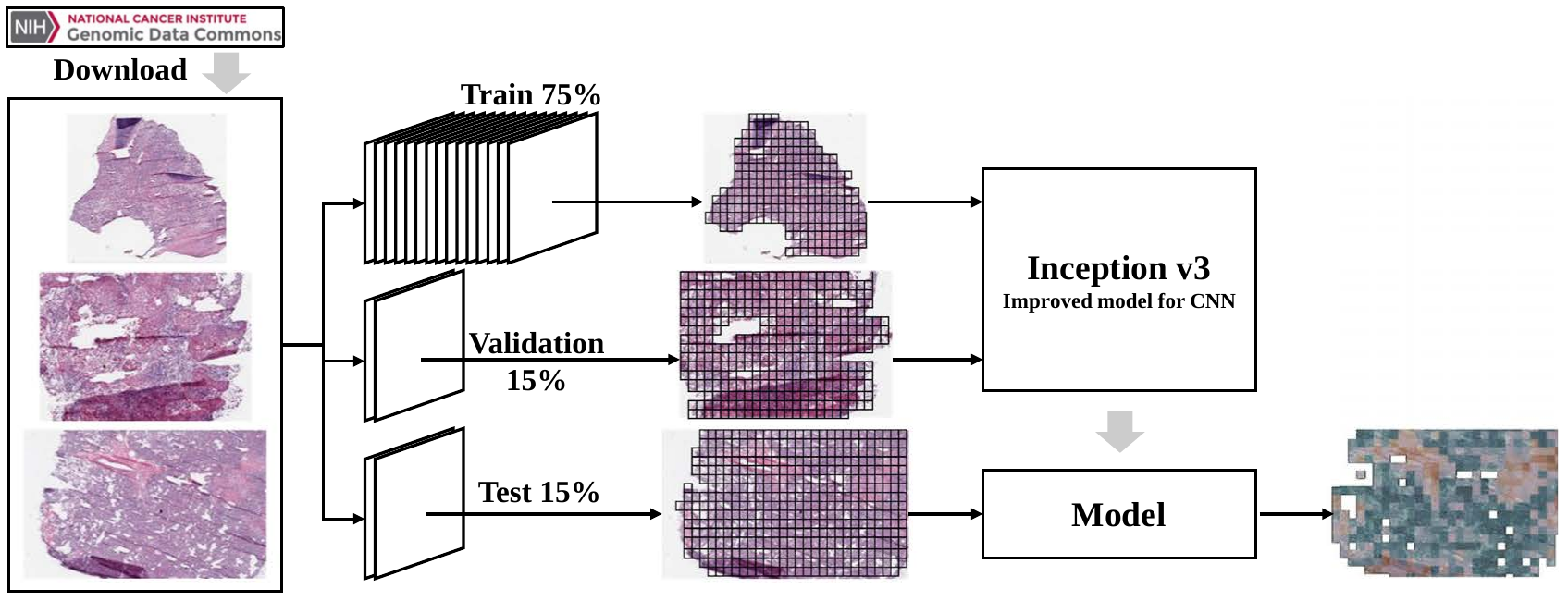,angle=0, width=3.25in}
\end{center}
\caption{The pipeline of work from Coudray et al.~\cite{coudray2018classification} which mainly including data preprocessing and model training by CNN.}%\vspace{-3mm}
\label{P1}
\end{figure}

In segmentation, Coudray et al.~\cite{coudray2018classification} show an excellent work on whole-slide images (obtained from The Cancer Genome Atlas, TCGA) to classify the Adenocarcinoma (LUAD), Squamous Cell Carcinoma (LUSC), and healthy lung tissue using Inception v3 (improved model of CNN~\cite{szegedy2016rethinking}). Figure~\ref{P1} is the structure of this work. Also, the author validated their work on independent datasets of frozen tissues, formalin-fixed paraffin-embedded tissues, and biopsies obtained at the New York University (NYU) Langone Medical Center. Furthermore, the trained network in this work could predict gene mutations using images as the only input in LUAD, which is meant to assist pathologists in the detection of cancer subtype or gene mutations. The Average area Under the Curve (AUC) of classification is 0.97, and AUCs of prediction is from 0.733 to 0.856.

\begin{figure}[!t]
\begin{center}
      \epsfig{figure=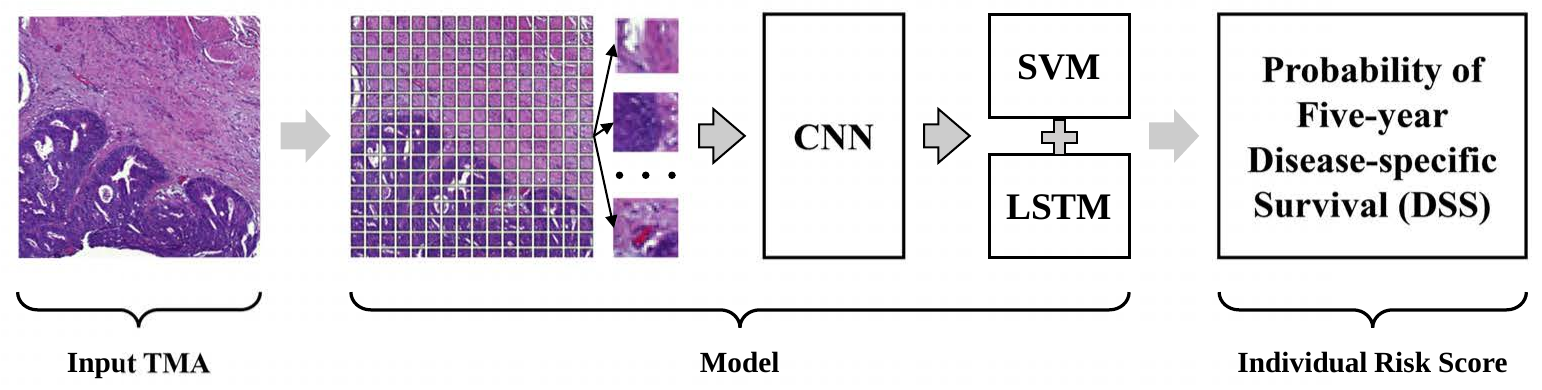,angle=0, width=3.25in}
\end{center}
\caption{The pipeline of the work written by Bychkov et al.~\cite{bychkov2018deep} which using CNN and LSTM to predict survival rate.}%\vspace{-3mm}
\label{P3}
\end{figure}

In prediction, Bychkov et al.~\cite{bychkov2018deep} combines CNN and Long Short-Term Memory (LSTM~\cite{hochreiter1997long}) to predict colorectal cancer outcome based on images of Haematoxylin and Eosin (H \& E) stained Tumour Tissue Microarray (TMA), which was shown in Figure~\ref{P3}. Also, this work can directly predict five-year disease-specific survival without any intermediate tissue classification. This work shows that extract more prognostic information from the tissue morphology of colorectal cancer than an experienced human observer.

As a summery, histology images in clinical practice are mainly used for tumor staging, grading, and cancer diagnosis. Therefore, classification is the priority in computer-assisted pathological diagnosis. In contrast, segmentation is about better identification of target cells, whereas prediction is about broader classification (predicting the likely consequences of future tumors). Therefore, it seems that there is no room for further development in the application of histology images. However, caused by the large capacity of pathological images, the current mainstream processing methods are training by blocks. It means that newer and better deep learning methods may be developed in the field to improve training speed or realize end-to-end.

\subsection{Treatment Planning}

After the diagnosis in the first three stages, doctors set out Treatment Plans according to the obtained diagnosis results, including operation therapy, radiotherapy, chemical therapy, and targeted molecular therapy. Operation therapy and radiotherapy belong to local treatment, which requires a clear division of Gross Tumor Volume (GTV), Clinical Tumor Volume (CTV), and Plan Tumor Volume (PTV) at the leisure. Therefore, images of tumors will be a suitable medium for operation therapy and radiotherapy. Chemotherapy is systemic therapy, which evaluated according to the improvement of clinical symptoms and objective indicators such as medical imaging. Molecular targeted therapy based on the level of cell molecules and therapeutic drug designs for oncogenic sites (such as protein molecules or gene fragments inside tumor cells), which can lead to specific death of tumor cells. Therefore, the data type in Treatment Planning is still image and gene expression by arrays.

Now, in the stage of treatment planning, doctors have demands on accurate lesion segmentation and reliable prediction of toxicity, survival, and efficacy, the details shown in Table~\ref{TP}.

\begin{table}[ht]
  \caption{Recent work in Treatment Planning}
  \centering
  \label{TP}
  \setlength{\tabcolsep}{3mm}{
  \begin{tabular}{c|c|c}
  \hline
  {Data} & {Tasks}  & {Reference} \\
  \hline
  {MRI} & \multirow{6}{*}{Prediction} & {\cite{li2017deep}, \cite{kao2018brain}, \cite{isensee2017brain}}\\
  \cline{1-1}
  \cline{3-3}
  {CT} & & \tabincell{c}{\cite{jackson2018deep}, \cite{foote2018real}, \cite{oakden2017precision}, \cite{wang2019deep},\\ \cite{zhao20183d}, \cite{ardila2019end}}\\
  \cline{1-1}
  \cline{3-3}
  {PET-CT} & & {\cite{zhang2017personalized}}\\
  \cline{1-1}
  \cline{3-3}
  \tabincell{c}{Histology\\images} & & {\cite{bychkov2018deep}, \cite{schaumberg2018h}, \cite{zhao20183d}}\\
  \cline{1-1}
  \cline{3-3}
  {Gene} & & {\cite{chang2018cancer}, \cite{xiao2018deep}, \cite{chaudhary2018deep}}\\
  \cline{1-1}
  \cline{3-3}
  \tabincell{c}{Treatment\\plans} & & \tabincell{c}{\cite{nguyen2017dose}, \cite{nguyen2018three}, \cite{nguyen2019feasibility},\\ \cite{nguyen20193d} \cite{zhen2017deep}}\\
  \hline
  \end{tabular}}
\end{table}

The prediction of the tumor through MRI, CT, PET-CT, and histology images has been introduced a lot before; next, the researches about genomic data, treatment plans, Rectum Surface Dose Maps (RSDM) will be analyzed mainly.

\begin{figure}[!t]
\begin{center}
      \epsfig{figure=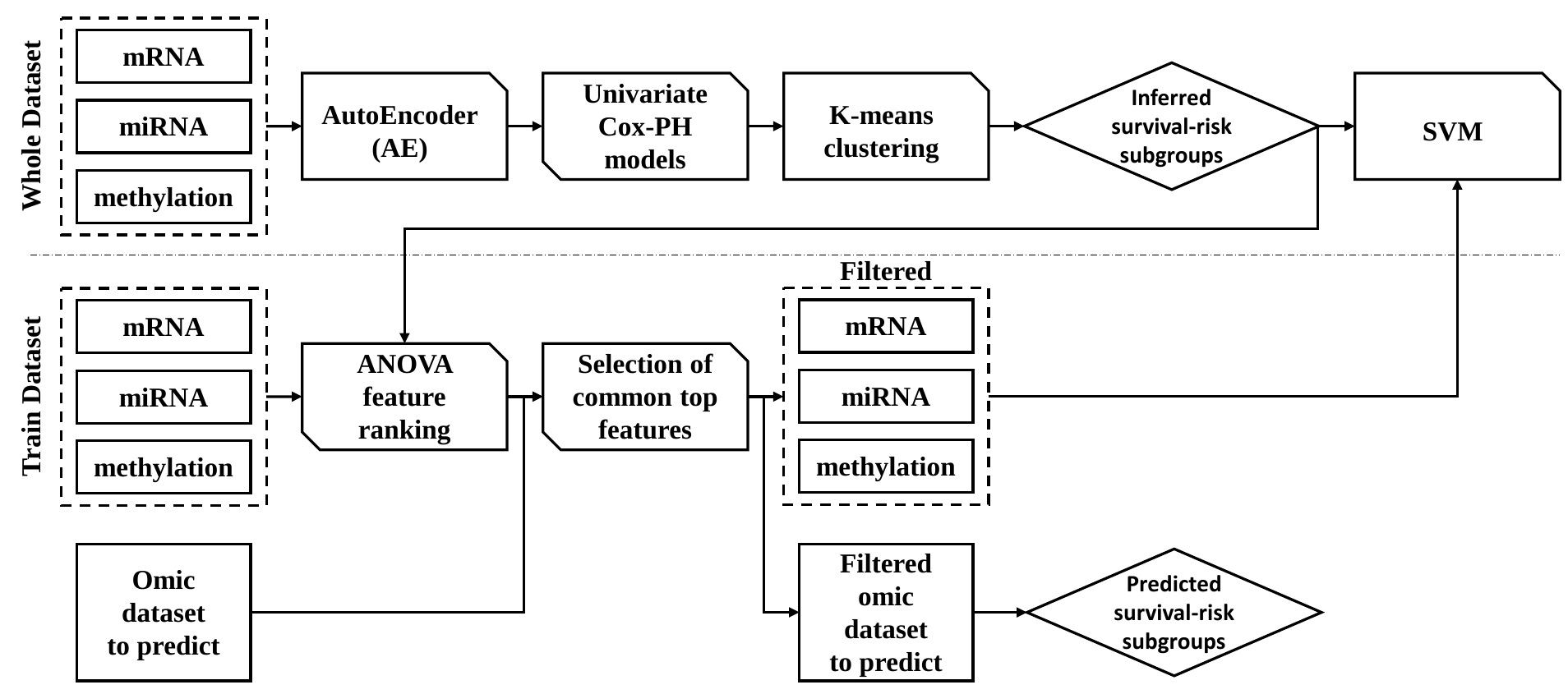,angle=0, width=3.25in}
\end{center}
\caption{The pipeline of the work from Chaudhary et al.~\cite{chaudhary2018deep}, an AutoEncoder architecture used to integrate 3 omics of HCC data.}%\vspace{-3mm}
\label{T1}
\end{figure}

In genomic data, Chaudhary et al.~\cite{chaudhary2018deep} uses a synthesis method combined by AutoEncoder (AE~\cite{hinton2006reducing}), K-means clustering~\cite{hartigan1979algorithm}, and Support Vector Machine (SVM~\cite{burges1998tutorial}), see Figure~\ref{T1}, to predict Hepatocellular Carcinoma (HCC) survival. The data in this work contains RNA sequencing (RNA-Seq), miRNA sequencing (miRNA-Seq), and methylation data from The Cancer Genome Atlas (TCGA). This effort predicts prognosis as good as an alternative model where genomics and clinical data are both considered.

\begin{figure}[!t]
\begin{center}
      \epsfig{figure=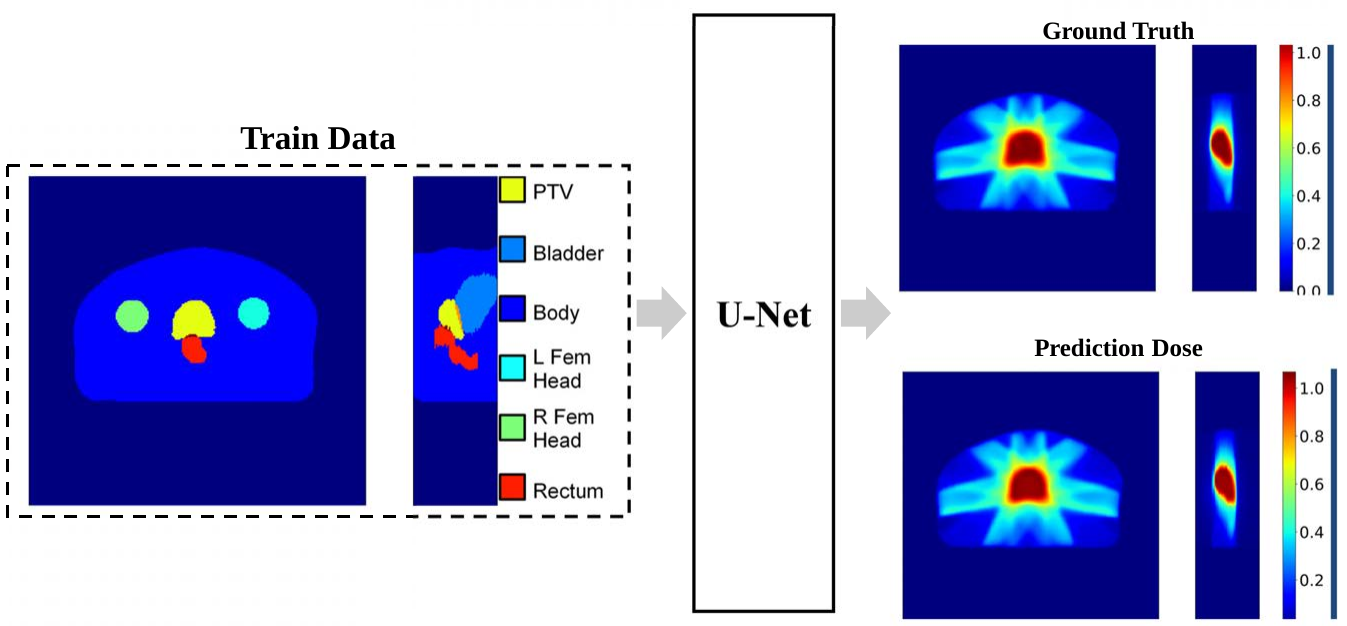,angle=0, width=3.25in}
\end{center}
\caption{The piplilne of treatment planning advanced by U-Net to predict the dose of radiotheropy~\cite{nguyen2019feasibility}.}%\vspace{-3mm}
\label{T2}
\end{figure}

In treatment plans, Nguyen et al. focus on predicting the dose distribution of radiotherapy, all the papers published by them~\cite{nguyen2017dose, nguyen2018three, nguyen2019feasibility, nguyen20193d} used U-Net to predict dose of radiotherapy. Such as the work~\cite{nguyen2019feasibility}, which shows in Figure~\ref{T2}, it is the first fully 3D dose distribution prediction for prostate IMRT plans. In this study, they use a similar set of 7 beam angles and criteria for treatment, which means that the model has currently learned only to predict the dose coming from approximately the same orientations, and may not be able to account for more intricate beam geometries. Also, the current model is unable to account for any physician preferences for predicting the dose, limiting the level of treatment personalization for the patient. However, the works of Nguyen et al. develop the dose prediction of cancer, which is meaningful.

As a summary, the works in this part are heavy and complex, which contains many kinds of prediction, but only a few of them are relevant. Although these efforts cover many areas (such as prediction of cancer~\cite{xiao2018deep}, life~\cite{oakden2017precision} and survival~\cite{kao2018brain, isensee2017brain}), few actually assist doctors in planning their treatment (such as prediction of radiation dose~\cite{jackson2018deep, nguyen2017dose, nguyen2018three, nguyen2019feasibility, nguyen20193d}, radiotoxicity~\cite{zhen2017deep}, anticancer drug response~\cite{chang2018cancer}, and recurrence forecast~\cite{wang2019deep}). There is still a dearth of real value studies on treatment plans, perhaps because it is difficult for non-medical researchers to collect data on treatment plans. Most studies only use some prediction as an ancillary result of research work, and there is still a huge gap in research that focuses on deep learning to assist tumor therapy. At this stage, the Computer-Aided Treatment of Tumors still needs more attention.

\section{Conclusion and Outlook}
\label{sec:Conclusionandoutlook}
\subsection{Conclusion}

In conclusion, how to achieve early detection of tumors, accurate diagnosis, proper treatment, and better prognosis becomes the key to tumor treatment. With the rapid development of deep learning, doctors increasingly need Computer-Aided Diagnosis and Treatment of Tumors, because the material defects of human beings limit the further development of this field. These defects usually caused by sensory thresholds, cognitive biases, and personal experience differences. In particular, deep learning has achieved great success in natural datasets to different task types. Specific to the data types (such as images, indicators and gene expression by arrays) and different medical tasks (such as classification, segmentation, detection, and prediction) in Computer-Aided Diagnosis and Treatment of Tumors, deep learning can provide effective methods (such as CNN, FCN, R-CNNs). These successes provide good reference cases and experience for a series of medical tasks of tumor diagnosis and treatment.

In order to entirely solve clinical problems by using deep learning requires a detailed understanding of the specific characteristics of the medical data and medical tasks in tumor diagnosis and treatment, which have been summarised as follows.

\begin{itemize}
\item Medical tasks focus on a single organ and kinds of leisure on the organ, such as the calcification, nodule, cyst, and tumor on the lung.
\item Medical tasks rely on stable human structures, such as the primary or metastatic lesions of the intracranial tumor within the skull.
\item The object in medical tasks is small and has individual differences, such as the pancreas and the thyroid.
\item Medical tasks are multidisciplinary collaborative diagnosis and treatment.
\end{itemize}

Compared with these research which published in top magazine~\cite{saltz2018spatial, kamnitsas2017efficient, zhao2018deep, drozdzal2018learning, klein2019maldi, chaudhary2018deep, coudray2018classification, falk2019u, ardila2019end}, it is easy to see, these works are either a meaningful breakthrough in the improvement of the new model (such as DeepMedic and U-Net) based on medical datasets or very close to the professional needs of medical tasks. Both types of studies have fully understood the clinical need and put it into practice. It means, when we are task-oriented and familiar with the characteristics of medical data, the deep learning method in computer science enables the rapid development of Computer-Aided Diagnosis and Treatment of Tumors. On the other hand, with the rapid development of deep learning in computer science, only by clarifying the nature and factions of various deep learning methods can researchers avoid detours.

This paper reviews the recent work in Computer-Aided Diagnosis and Treatment of Tumors from the following four points, data type, organs, medical tasks, and deep learning method. These works are of great significance in the Computer-Aided Diagnosis and Treatment of Tumors. The specific content is summarized as follows.

Firstly, the survey summarizes the data type of tumor diagnosis and treatment from four stages. Each stage may have a different data type depending on its characteristics or may have a different tendency to process the data according to the requirements of the medical task.
\begin{itemize}
\item Stage of In-Vitro Diagnosis (IVD). There are three kinds of medical data in this stage. 1) Images from the microscope, blood smear, and Flow Cytometer (FCM). 2) arrays of cancer-associated genomic regions. Also, 3) numerical indices from a Flow Cytometer (FCM).
\item Stage of Image Diagnosis. In this stage, most data is images from various medical imaging devices, such as Magnetic Resonance Imaging (MRI), Computed Tomography (CT), Single-PPhoton EEmission Computed Tomography (SPECT), Positron Emission Tomography (PET), Ultrasound (US), mammography.
\item Stage of Pathological Diagnosis. In this stage, histology images stained by H \& E is the mean medical data type.
\item Stage of Treatment Plan. In this stage, except for the images and genomic data summarised above, radiotherapy data such as rectum surface dose maps (RSDM) is meaningful.
\end{itemize}

Secondly, the article reviews majority of organs and tissues which easy to have tumors.
\begin{itemize}
\item The target organ of the tumor. Such as the brain, lung, colorectum, nasopharynx, liver, kidney, bladder, stomach, cervix, head \& neck.
\item The target gland of the tumor. Such as breast, prostate, thyroid, ovary, and pancreas.
\item The organ at risk (OAR). Such as the heart, esophagus, trachea, and aorta.
\item Connective tissue, such as blood.
\end{itemize}

Thirdly, this work contains several medical tasks in four stages of tumor diagnosis and treatment, details was shown as follows.
\begin{itemize}
\item Detection, common medical tasks include in this part are screening the tumor, the carcinoma cells, and the biomarkers.
\item Segmentation, such as delineating the lesion and organ at risk (OAR).
\item Classification, such as identifying the tumor subtypes and stages.
\item Prediction, which contains a lot of clinical practice types. Including 1) prediction of radiation dose, radiotoxicity, anticancer drug response, and recurrence forecast; 2) motion prediction under the action of breathing; 3) prediction of cancer, tumor invasiveness, tumor growth; and 4) prediction of life and survival.
\end{itemize}

Finally, the state-of-the-art in the Computer-Aided Diagnosis and Treatment of Tumors have been introduced detailedly before. It is a task-oriented review of different task types and their appropriate deep learning methods, such as the CNN series in the field of classification, the FCN series in the field of segmentation, and the R-CNN series of object detection. In general, these models are better at image data processing. However, recently researches are feeble in the area of non-image data which is also an essential source of clinical data.

Impressively, there are still plenty of reasons limit to the development of the Computer-Aided Diagnosis and Treatment of Tumors, which need the medical industry and the computer industry through cooperation can be targeted to solve. Such as medical imaging lacking high-quality datasets with unified standards and accurate labeling, existing models easy to over-fit with less robustness, data security, and diagnostic reliability. Therefore, there are still plenty of works is waiting for researchers.

\subsection{Outlook and Challenge}

According to the above summary and analysis, the Computer-Aided Diagnosis and Treatment of Tumors have thrived in all aspects of clinical practice. However, there are still many gaps and challenges worth noted based on this work. The following sections elaborate on these issues in four ways.

\subsubsection{From Data Types}

The data types in recent researches seem to be quite comprehensive, but more often than not, people tend to focus on the Image Diagnosis data. Even this tendency extends to a few kinds of medical images, such as MRI and CT. Histology images and cancer-associated genomic regions have been partially concerned. However, some problems limit the progress of the research. For individual histology images, it can be up to 2G in size, which requires advanced servers for researchers. Also, lots of primary hospitals have no conditions for genetic testing. Besides, the data from stages of In-Vitro Diagnosis (IVD) and Treatment Plan have the same predicaments - there is not exist a sophisticated system to store the data of tumor diagnosis and treatment in bulks.

Meanwhile, the multi-disciplinary collaboration of medical tasks makes the multi-type datasets of the same task have a strong correlation, and taking data groups as research objects will gradually become a trend. Such as the integration of genomics and histopathology mentioned in this review~\cite{coudray2018classification}.

Above all, deep learning is an art of big data, the gaps in early screening, treatment of tumors, and the study of multi-omics still need to be filled.

\subsubsection{From Organs and Tissues}

There are almost 20 different organs have been mentioned in this survey. However, the researchers prefer to focus on a few large organs (such as brain, liver, and lung), and the organ who has special examination (such as breast image from mammography, and thyroid image from ultrasound). Therefore, some structurally specific organ which is tiny and deep in humans body lacks the correlational research, such as the pancreas. The diagnosis and treatment of tumors in these organs is more challenging and meaningful,especially the pancreatic cancer is short disease duration and fatal. In this part, there is plenty of work which focus on tiny organs and tissues has to follow up.

\subsubsection{From Medical Tasks}

In this paper, more than 17 kinds of medical tasks in the Computer-Aided Diagnosis and Treatment of Tumors have been summarized. It is shown that the majority of the research priorities are detection, segmentation, and classification. There are screening the tumors, carcinoma cells, and biomarkers; delineating the lesion and Organ At Risk (OAR); and identifying the tumor subtypes and stages. However, in the prediction tasks of tumors, a unique phenomenon has emerged.
Interestingly, the vast majority of studies on prediction of survival, prediction of life span do not contribute to clinical practice; they are usually just additional contributions from other studies. Although it seems significant, this kind of research tells us that patients are more likely to die if they get malignant tumors. These works do not provide constructive Suggestions to help doctors plan their next treatment. On the contrary, studies on actual adjuvant therapy, such as exercise management, radiation dose distribution prediction, radiation toxicity prediction, drug effect prediction, and tumor erosion prediction, are of clinical significance. However, due to the scarcity of data sources, this kind of research progress is languid, and more researchers need to put in relevant work.

\subsubsection{From Deep Learning Methods}

Although the application in the Computer-Aided Diagnosis and Treatment of Tumors of various deep learning models is generally in line with the development trend of computer science, the application is still sluggish. At present, the research in the field is still limited to simple deep learning model handling, and U-Net is the only influential model improved by medical data. There are gaps that state-of-the-art models published recently are not yet widely used in the Computer-Aided Diagnosis and Treatment of Tumors, such as \textit{Segmentation is all you need}~\cite{wu2019segmentation}, a new revolution in object detection which good at difficulty tiny object; Transformer~\cite{vaswani2017attention}, BERT~\cite{devlin2018bert}, Transformer-XL~\cite{dai2019transformer}, XLNet~\cite{yang2019xlnet}, Sparse Transformers~\cite{child2019generating} in Natural Language Processing (NLP), which could be used for clinical data, such as medical history. It may revolutionize the performance of current oncology adjuvant therapy if the researchers introduce these advanced methods. Of course, it is particularly welcome if researchers are working on improving deep or creating learning models based on medical data and aim at medical tasks.

\subsubsection{Adversarial Study}

Given the existing problems in tumor diagnosis and treatment in the whole industry, this paper has made a prospect and put forward the vacancy of existing research. However, in essence, some critical studies have never been done before, and it is worth noting. Adversarial study is one of such field.

Any research in deep learning is bound to encounter the question of whether the results are credible. In such an information age, information security and ethics are worth discussing. It remains to be seen whether researchers will be able to achieve consistent results in simple information confrontation experiments in the field of tumor diagnosis and treatment, not to mention that the diagnosis and treatment of tumors are vital and should be treated with caution. As far as we know, this field is almost blank except the recent work~\cite{finlayson2019adversarial}, which is a very terrible blank, but also a blank with infinite opportunities.

As a summary, Computer-Aided Diagnosis and Treatment of Tumors are highly cross-disciplinary that ask researchers know medicine and computer science, but few people are proficient in both, and not everyone can find a partner with expertise. This problem has led to a situation in the field where computer scientists do not know how to make medical advances in combination with specific conditions, and doctors do not know which deep learning methods can better achieve their goals. It is also the reason why a large number of studies have been conducted which cannot guide clinical tumor treatment. For now, this paper could quickly guide doctors to choose a proper deep learning method, and computer researchers may learn more about tumor diagnosis and treatment. In clinical practice, the international standard for tumor diagnosis and treatment refer to the National Comprehensive Cancer Network (NCCN) Clinical Practice Guidelines in Oncology. (see: \textbf{\url{https://www.nccn.org/professionals/physician_gls/default.aspx}}). If researchers want to further improve the fit between their works and the medical task, clicking on it would be a good start.

\section*{Acknowledgments}
\label{sec:Acknowledgements}
\noindent This work was supported in part by the National Natural Science Foundation of China under grant 61906063, in part by the Natural Science Foundation of Tianjin, China, under grant 19JCQNJC00400, and in part by the Yuanguang Scholar Fund of Hebei University of Technology, China. 

%\input{RelatedWork}
%\input{Preliminary}
%\input{ELSA}
%\input{LP-ESA}
%\input{LP-DSA}
%\input{Experiments}
%\input{Conclusion}
%\section*{References}
\bibliographystyle{elsarticle-num}
\bibliography{elsarticle-template_zd.bbl}

\begin{thebibliography}{100}
\expandafter\ifx\csname url\endcsname\relax
  \def\url#1{\texttt{#1}}\fi
\expandafter\ifx\csname urlprefix\endcsname\relax\def\urlprefix{URL }\fi
\expandafter\ifx\csname href\endcsname\relax
  \def\href#1#2{#2} \def\path#1{#1}\fi

\bibitem{bray2018global}
F.~Bray, J.~Ferlay, I.~Soerjomataram, R.~L. Siegel, L.~A. Torre, A.~Jemal,
  Global cancer statistics 2018: Globocan estimates of incidence and mortality
  worldwide for 36 cancers in 185 countries, CA: A Cancer Journal for
  Clinicians 68~(6) (2018) 394--424.

\bibitem{hinton2006reducing}
G.~E. Hinton, R.~R. Salakhutdinov, Reducing the dimensionality of data with
  neural networks, Science 313~(5786) (2006) 504--507.

\bibitem{lecun1998gradient}
Y.~LeCun, L.~Bottou, Y.~Bengio, P.~Haffner, et~al., Gradient-based learning
  applied to document recognition, Proceedings of the IEEE 86~(11) (1998)
  2278--2324.

\bibitem{girshick2014rich}
R.~Girshick, J.~Donahue, T.~Darrell, J.~Malik, Rich feature hierarchies for
  accurate object detection and semantic segmentation, in: In Proceedings of
  the IEEE Conference on Computer Vision and Pattern Recognition, 2014, pp.
  580--587.

\bibitem{girshick2015fast}
R.~Girshick, Fast r-cnn, in: In Proceedings of the IEEE International
  Conference on Computer Vision, 2015, pp. 1440--1448.

\bibitem{ren2015faster}
S.~Ren, K.~He, R.~Girshick, J.~Sun, Faster r-cnn: Towards real-time object
  detection with region proposal networks, in: Advances in Neural Information
  Processing Systems, 2015, pp. 91--99.

\bibitem{long2015fully}
J.~Long, E.~Shelhamer, T.~Darrell, Fully convolutional networks for semantic
  segmentation, in: In Proceedings of the IEEE Conference on Computer Vision
  and Pattern Recognition, 2015, pp. 3431--3440.

\bibitem{ronneberger2015u}
O.~Ronneberger, P.~Fischer, T.~Brox, U-net: Convolutional networks for
  biomedical image segmentation, in: In Proceedings of the International
  Conference on Medical Image Computing and Computer-Assisted Intervention,
  2015, pp. 234--241.

\bibitem{lipton2015critical}
Z.~C. Lipton, J.~Berkowitz, C.~Elkan, A critical review of recurrent neural
  networks for sequence learning, ArXiv Preprint ArXiv:1506.00019.

\bibitem{hochreiter1997long}
S.~Hochreiter, J.~Schmidhuber, Long short-term memory, Neural Computation 9~(8)
  (1997) 1735--1780.

\bibitem{ker2017deep}
J.~Ker, L.~Wang, J.~Rao, T.~Lim, Deep learning applications in medical image
  analysis, IEEE Access 6 (2017) 9375--9389.

\bibitem{meyer2018survey}
P.~Meyer, V.~Noblet, C.~Mazzara, A.~Lallement, Survey on deep learning for
  radiotherapy, Computers in Biology and Medicine 98 (2018) 126--146.

\bibitem{yasaka2018deep}
K.~Yasaka, H.~Akai, A.~Kunimatsu, S.~Kiryu, O.~Abe, Deep learning with
  convolutional neural network in radiology, Japanese Journal of Radiology
  36~(4) (2018) 257--272.

\bibitem{sahiner2019deep}
B.~Sahiner, A.~Pezeshk, L.~M. Hadjiiski, X.~Wang, K.~Drukker, K.~H. Cha, R.~M.
  Summers, M.~L. Giger, Deep learning in medical imaging and radiation therapy,
  Medical Physics 46~(1) (2019) e1--e36.

\bibitem{hu2018deep}
Z.~Hu, J.~Tang, Z.~Wang, K.~Zhang, L.~Zhang, Q.~Sun, Deep learning for
  image-based cancer detection and diagnosis- a survey, Pattern Recognition 83
  (2018) 134--149.

\bibitem{ueda2019technical}
D.~Ueda, A.~Shimazaki, Y.~Miki, Technical and clinical overview of deep
  learning in radiology, Japanese Journal of Radiology 37~(1) (2019) 15--33.

\bibitem{liu2018applications}
J.~Liu, Y.~Pan, M.~Li, Z.~Chen, L.~Tang, C.~Lu, J.~Wang, Applications of deep
  learning to mri images: A survey, Big Data Mining and Analytics 1~(1) (2018)
  1--18.

\bibitem{mazurowski2019deep}
M.~A. Mazurowski, M.~Buda, A.~Saha, M.~R. Bashir, Deep learning in radiology:
  An overview of the concepts and a survey of the state of the art with focus
  on mri, Journal of Magnetic Resonance Imaging 49~(4) (2019) 939--954.

\bibitem{liu2019deep}
S.~Liu, Y.~Wang, X.~Yang, B.~Lei, L.~Liu, S.~X. Li, D.~Ni, T.~Wang, Deep
  learning in medical ultrasound analysis: A review, Engineering 5~(2) (2019)
  261--275.

\bibitem{napel2018quantitative}
S.~Napel, W.~Mu, B.~V. Jardim-Perassi, H.~J. Aerts, R.~J. Gillies, Quantitative
  imaging of cancer in the postgenomic era: Radio (geno) mics, deep learning,
  and habitats, Cancer 124~(24) (2018) 4633--4649.

\bibitem{cao2018deep}
C.~Cao, F.~Liu, H.~Tan, D.~Song, W.~Shu, W.~Li, Y.~Zhou, X.~Bo, Z.~Xie, Deep
  learning and its applications in biomedicine, Genomics, Proteomics \&
  Bioinformatics 16~(1) (2018) 17--32.

\bibitem{shen2017deep}
D.~Shen, G.~Wu, H.-I. Suk, Deep learning in medical image analysis, Annual
  Review of Biomedical Engineering 19 (2017) 221--248.

\bibitem{razzak2018deep}
M.~I. Razzak, S.~Naz, A.~Zaib, Deep learning for medical image processing:
  Overview, challenges and the future, in: Classification in BioApps, 2018, pp.
  323--350.

\bibitem{ching2018opportunities}
T.~Ching, D.~S. Himmelstein, B.~K. Beaulieu-Jones, A.~A. Kalinin, B.~T. Do,
  G.~P. Way, E.~Ferrero, P.-M. Agapow, M.~Zietz, M.~M. Hoffman, et~al.,
  Opportunities and obstacles for deep learning in biology and medicine,
  Journal of The Royal Society Interface 15~(141) (2018) 20170387.

\bibitem{akkus2019survey}
Z.~Akkus, J.~Cai, A.~Boonrod, A.~Zeinoddini, A.~D. Weston, K.~A. Philbrick,
  B.~J. Erickson, A survey of deep-learning applications in ultrasound:
  Artificial intelligence--powered ultrasound for improving clinical workflow,
  Journal of the American College of Radiology 16~(9) (2019) 1318--1328.

\bibitem{finlayson2019adversarial}
S.~G. Finlayson, J.~D. Bowers, J.~Ito, J.~L. Zittrain, A.~L. Beam, I.~S.
  Kohane, Adversarial attacks on medical machine learning, Science 363~(6433)
  (2019) 1287--1289.

\bibitem{he2017mask}
K.~He, G.~Gkioxari, P.~Doll{\'a}r, R.~Girshick, Mask r-cnn, in: In Proceedings
  of the IEEE international conference on computer vision, 2017, pp.
  2961--2969.

\bibitem{vaswani2017attention}
A.~Vaswani, N.~Shazeer, N.~Parmar, J.~Uszkoreit, L.~Jones, A.~N. Gomez,
  {\L}.~Kaiser, I.~Polosukhin, Attention is all you need, in: Advances in
  neural information processing systems, 2017, pp. 5998--6008.

\bibitem{kamnitsas2017efficient}
K.~Kamnitsas, C.~Ledig, V.~F. Newcombe, J.~P. Simpson, A.~D. Kane, D.~K. Menon,
  D.~Rueckert, B.~Glocker, Efficient multi-scale 3d cnn with fully connected
  crf for accurate brain lesion segmentation, Medical Image Analysis 36 (2017)
  61--78.

\bibitem{simonyan2014very}
K.~Simonyan, A.~Zisserman, Very deep convolutional networks for large-scale
  image recognition, ArXiv Preprint ArXiv:1409.1556.

\bibitem{he2016deep}
K.~He, X.~Zhang, S.~Ren, J.~Sun, Deep residual learning for image recognition,
  in: In Proceedings of the IEEE conference on computer vision and pattern
  recognition, 2016, pp. 770--778.

\bibitem{szegedy2015going}
C.~Szegedy, W.~Liu, Y.~Jia, P.~Sermanet, S.~Reed, D.~Anguelov, D.~Erhan,
  V.~Vanhoucke, A.~Rabinovich, Going deeper with convolutions, in: In
  Proceedings of the IEEE conference on computer vision and pattern
  recognition, 2015, pp. 1--9.

\bibitem{bellver2017detection}
M.~Bellver, K.-K. Maninis, J.~Pont-Tuset, X.~Gir{\'o}-i Nieto, J.~Torres,
  L.~Van~Gool, Detection-aided liver lesion segmentation using deep learning,
  ArXiv Preprint ArXiv:1711.11069.

\bibitem{milletari2017hough}
F.~Milletari, S.-A. Ahmadi, C.~Kroll, A.~Plate, V.~Rozanski, J.~Maiostre,
  J.~Levin, O.~Dietrich, B.~Ertl-Wagner, K.~B{\"o}tzel, et~al., Hough-cnn: deep
  learning for segmentation of deep brain regions in mri and ultrasound,
  Computer Vision and Image Understanding 164 (2017) 92--102.

\bibitem{araujo2017classification}
T.~Ara{\'u}jo, G.~Aresta, E.~Castro, J.~Rouco, P.~Aguiar, C.~Eloy,
  A.~Pol{\'o}nia, A.~Campilho, Classification of breast cancer histology images
  using convolutional neural networks, PloS One 12~(6) (2017) e0177544.

\bibitem{cruz2017accurate}
A.~Cruz-Roa, H.~Gilmore, A.~Basavanhally, M.~Feldman, S.~Ganesan, N.~N. Shih,
  J.~Tomaszewski, F.~A. Gonz{\'a}lez, A.~Madabhushi, Accurate and reproducible
  invasive breast cancer detection in whole-slide images: A deep learning
  approach for quantifying tumor extent, Scientific Reports 7 (2017) 46450.

\bibitem{saltz2018spatial}
J.~Saltz, R.~Gupta, L.~Hou, T.~Kurc, P.~Singh, V.~Nguyen, D.~Samaras, K.~R.
  Shroyer, T.~Zhao, R.~Batiste, et~al., Spatial organization and molecular
  correlation of tumor-infiltrating lymphocytes using deep learning on
  pathology images, Cell Reports 23~(1) (2018) 181--193.

\bibitem{golatkar2018classification}
A.~Golatkar, D.~Anand, A.~Sethi, Classification of breast cancer histology
  using deep learning, in: In Proceedings of the International Conference on
  Image Analysis and Recognition, 2018, pp. 837--844.

\bibitem{khoshdeli2018deep}
M.~Khoshdeli, A.~Borowsky, B.~Parvin, Deep learning models differentiate tumor
  grades from h\&e stained histology sections, in: In Proceedings of the IEEE
  Engineering in Medicine and Biology Society, 2018, pp. 620--623.

\bibitem{khosravi2018deep}
P.~Khosravi, E.~Kazemi, M.~Imielinski, O.~Elemento, I.~Hajirasouliha, Deep
  convolutional neural networks enable discrimination of heterogeneous digital
  pathology images, EBioMedicine 27 (2018) 317--328.

\bibitem{saouli2018fully}
R.~Saouli, M.~Akil, R.~Kachouri, et~al., Fully automatic brain tumor
  segmentation using end-to-end incremental deep neural networks in mri images,
  Computer Methods and Programs in Biomedicine 166 (2018) 39--49.

\bibitem{li2017deep}
Z.~Li, Y.~Wang, J.~Yu, Y.~Guo, W.~Cao, Deep learning based radiomics (dlr) and
  its usage in noninvasive idh1 prediction for low grade glioma, Scientific
  Reports 7~(1) (2017) 5467.

\bibitem{laukamp2019fully}
K.~R. Laukamp, F.~Thiele, G.~Shakirin, D.~Zopfs, A.~Faymonville, M.~Timmer,
  D.~Maintz, M.~Perkuhn, J.~Borggrefe, Fully automated detection and
  segmentation of meningiomas using deep learning on routine multiparametric
  mri, European Radiology 29~(1) (2019) 124--132.

\bibitem{leung2018deep}
K.~Leung, W.~Marashdeh, R.~Wray, S.~Ashrafinia, A.~Rahmim, M.~Pomper, A.~Jha, A
  deep-learning-based fully automated segmentation approach to delineate tumors
  in fdg-pet images of patients with lung cancer, Journal of Nuclear Medicine
  59~(supplement 1) (2018) 323--323.

\bibitem{trebeschi2017deep}
S.~Trebeschi, J.~J. van Griethuysen, D.~M. Lambregts, M.~J. Lahaye, C.~Parmar,
  F.~C. Bakers, N.~H. Peters, R.~G. Beets-Tan, H.~J. Aerts, Deep learning for
  fully-automated localization and segmentation of rectal cancer on
  multiparametric mr, Scientific Reports 7~(1) (2017) 5301.

\bibitem{li2018tumor}
Q.~Li, Y.~Xu, Z.~Chen, D.~Liu, S.-T. Feng, M.~Law, Y.~Ye, B.~Huang, Tumor
  segmentation in contrast-enhanced magnetic resonance imaging for
  nasopharyngeal carcinoma: Deep learning with convolutional neural network,
  BioMed Research International 2018 (2018) 9128527.

\bibitem{janowczyk2018resolution}
A.~Janowczyk, S.~Doyle, H.~Gilmore, A.~Madabhushi, A resolution adaptive deep
  hierarchical (radhical) learning scheme applied to nuclear segmentation of
  digital pathology images, Computer Methods in Biomechanics and Biomedical
  Engineering: Imaging \& Visualization 6~(3) (2018) 270--276.

\bibitem{jackson2018deep}
P.~Jackson, N.~Hardcastle, N.~Dawe, T.~Kron, M.~Hofman, R.~J. Hicks, Deep
  learning renal segmentation for fully automated radiation dose estimation in
  unsealed source therapy, Frontiers in Oncology 8 (2018) 215.

\bibitem{foote2018real}
M.~D. Foote, B.~Zimmerman, A.~Sawant, S.~Joshi, Real-time patient-specific lung
  radiotherapy targeting using deep learning, ArXiv Preprint ArXiv:1807.08388.

\bibitem{zhen2017deep}
X.~Zhen, J.~Chen, Z.~Zhong, B.~Hrycushko, L.~Zhou, S.~Jiang, K.~Albuquerque,
  X.~Gu, Deep convolutional neural network with transfer learning for rectum
  toxicity prediction in cervical cancer radiotherapy: a feasibility study,
  Physics in Medicine \& Biology 62~(21) (2017) 8246.

\bibitem{oakden2017precision}
L.~Oakden-Rayner, G.~Carneiro, T.~Bessen, J.~C. Nascimento, A.~P. Bradley,
  L.~J. Palmer, Precision radiology: predicting longevity using feature
  engineering and deep learning methods in a radiomics framework, Scientific
  Reports 7~(1) (2017) 1648.

\bibitem{zhang2017personalized}
L.~Zhang, L.~Lu, R.~M. Summers, E.~Kebebew, J.~Yao, Personalized pancreatic
  tumor growth prediction via group learning, in: In Proceedings of the
  International Conference on Medical Image Computing and Computer-Assisted
  Intervention, 2017, pp. 424--432.

\bibitem{bychkov2018deep}
D.~Bychkov, N.~Linder, R.~Turkki, S.~Nordling, P.~E. Kovanen, C.~Verrill,
  M.~Walliander, M.~Lundin, C.~Haglund, J.~Lundin, Deep learning based tissue
  analysis predicts outcome in colorectal cancer, Scientific Reports 8~(1)
  (2018) 3395.

\bibitem{wang2019deep}
S.~Wang, Z.~Liu, Y.~Rong, B.~Zhou, Y.~Bai, W.~Wei, M.~Wang, Y.~Guo, J.~Tian,
  Deep learning provides a new computed tomography-based prognostic biomarker
  for recurrence prediction in high-grade serous ovarian cancer, Radiotherapy
  and Oncology 132 (2019) 171--177.

\bibitem{zhao20183d}
W.~Zhao, J.~Yang, Y.~Sun, C.~Li, W.~Wu, L.~Jin, Z.~Yang, B.~Ni, P.~Gao,
  P.~Wang, et~al., 3d deep learning from ct scans predicts tumor invasiveness
  of subcentimeter pulmonary adenocarcinomas, Cancer Research 78~(24) (2018)
  6881--6889.

\bibitem{bulten2019epithelium}
W.~Bulten, P.~B{\'a}ndi, J.~Hoven, R.~van~de Loo, J.~Lotz, N.~Weiss, J.~van~der
  Laak, B.~van Ginneken, C.~Hulsbergen-van~de Kaa, G.~Litjens, Epithelium
  segmentation using deep learning in h\&e-stained prostate specimens with
  immunohistochemistry as reference standard, Scientific Reports 9~(1) (2019)
  864.

\bibitem{perkuhn2018clinical}
M.~Perkuhn, P.~Stavrinou, F.~Thiele, G.~Shakirin, M.~Mohan, D.~Garmpis,
  C.~Kabbasch, J.~Borggrefe, Clinical evaluation of a multiparametric deep
  learning model for glioblastoma segmentation using heterogeneous magnetic
  resonance imaging data from clinical routine, Investigative Radiology 53~(11)
  (2018) 647--654.

\bibitem{kao2018brain}
P.-Y. Kao, T.~Ngo, A.~Zhang, J.~W. Chen, B.~Manjunath, Brain tumor segmentation
  and tractographic feature extraction from structural mr images for overall
  survival prediction, in: In Proceedings of the International Conference on
  Medical Image Computing and Computer Assisted Intervention Brainlesion
  Workshop, 2018, pp. 128--141.

\bibitem{kamnitsas2017ensembles}
K.~Kamnitsas, W.~Bai, E.~Ferrante, S.~McDonagh, M.~Sinclair, N.~Pawlowski,
  M.~Rajchl, M.~Lee, B.~Kainz, D.~Rueckert, et~al., Ensembles of multiple
  models and architectures for robust brain tumour segmentation, in: In
  Proceedings of the International Conference on Medical Image Computing and
  Computer Assisted Intervention Brainlesion Workshop, 2017, pp. 450--462.

\bibitem{castillo2017volumetric}
L.~S. Castillo, L.~A. Daza, L.~C. Rivera, P.~Arbel{\'a}ez, Volumetric
  multimodality neural network for brain tumor segmentation, in: In Proceedings
  of the International Conference on Medical Information Processing and
  Analysis, Vol. 10572, 2017, p. 105720E.

\bibitem{trullo2017segmentation}
R.~Trullo, C.~Petitjean, S.~Ruan, B.~Dubray, D.~Nie, D.~Shen, Segmentation of
  organs at risk in thoracic ct images using a sharpmask architecture and
  conditional random fields, in: In Proceedings of the IEEE International
  Symposium on Biomedical Imaging, 2017, pp. 1003--1006.

\bibitem{shen2017boundary}
H.~Shen, R.~Wang, J.~Zhang, S.~J. McKenna, Boundary-aware fully convolutional
  network for brain tumor segmentation, in: In Proceedings of the International
  Conference on Medical Image Computing and Computer-Assisted Intervention,
  2017, pp. 433--441.

\bibitem{zhao2018deep}
X.~Zhao, Y.~Wu, G.~Song, Z.~Li, Y.~Zhang, Y.~Fan, A deep learning model
  integrating fcnns and crfs for brain tumor segmentation, Medical Image
  Analysis 43 (2018) 98--111.

\bibitem{christ2017automatic}
P.~F. Christ, F.~Ettlinger, F.~Gr{\"u}n, M.~E.~A. Elshaera, J.~Lipkova,
  S.~Schlecht, F.~Ahmaddy, S.~Tatavarty, M.~Bickel, P.~Bilic, et~al., Automatic
  liver and tumor segmentation of ct and mri volumes using cascaded fully
  convolutional neural networks, ArXiv Preprint ArXiv:1702.05970.

\bibitem{soomro2018automatic}
M.~H. Soomro, G.~De~Cola, S.~Conforto, M.~Schmid, G.~Giunta, E.~Guidi, E.~Neri,
  D.~Caruso, M.~Ciolina, A.~Laghi, Automatic segmentation of colorectal cancer
  in 3d mri by combining deep learning and 3d level-set algorithm-a preliminary
  study, in: In Proceedings of the IEEE Middle East Conference on Biomedical
  Engineering, 2018, pp. 198--203.

\bibitem{drozdzal2018learning}
M.~Drozdzal, G.~Chartrand, E.~Vorontsov, M.~Shakeri, L.~Di~Jorio, A.~Tang,
  A.~Romero, Y.~Bengio, C.~Pal, S.~Kadoury, Learning normalized inputs for
  iterative estimation in medical image segmentation, Medical Image Analysis 44
  (2018) 1--13.

\bibitem{falk2019u}
T.~Falk, D.~Mai, R.~Bensch, {\"O}.~{\c{C}}i{\c{c}}ek, A.~Abdulkadir,
  Y.~Marrakchi, A.~B{\"o}hm, J.~Deubner, Z.~J{\"a}ckel, K.~Seiwald, et~al.,
  U-net: deep learning for cell counting, detection, and morphometry, Nature
  Methods 16~(1) (2019) 67.

\bibitem{beers2017sequential}
A.~Beers, K.~Chang, J.~Brown, E.~Sartor, C.~Mammen, E.~Gerstner, B.~Rosen,
  J.~Kalpathy-Cramer, Sequential 3d u-nets for biologically-informed brain
  tumor segmentation, ArXiv Preprint ArXiv:1709.02967.

\bibitem{isensee2017brain}
F.~Isensee, P.~Kickingereder, W.~Wick, M.~Bendszus, K.~H. Maier-Hein, Brain
  tumor segmentation and radiomics survival prediction: Contribution to the
  brats 2017 challenge, in: In Proceedings of the International Conference on
  Medical Image Computing and Computer Assisted Intervention Brainlesion
  Workshop, 2017, pp. 287--297.

\bibitem{guo2018deep}
Z.~Guo, L.~Zhang, L.~Lu, M.~Bagheri, R.~M. Summers, M.~Sonka, J.~Yao, Deep
  logismos: Deep learning graph-based 3d segmentation of pancreatic tumors on
  ct scans, in: In Proceedings of the IEEE International Symposium on
  Biomedical Imaging, 2018, pp. 1230--1233.

\bibitem{zhong20183d}
Z.~Zhong, Y.~Kim, L.~Zhou, K.~Plichta, B.~Allen, J.~Buatti, X.~Wu, 3d fully
  convolutional networks for co-segmentation of tumors on pet-ct images, in: In
  Proceedings of the IEEE International Symposium on Biomedical Imaging, 2018,
  pp. 228--231.

\bibitem{nguyen2017dose}
D.~Nguyen, T.~Long, X.~Jia, W.~Lu, X.~Gu, Z.~Iqbal, S.~Jiang, Dose prediction
  with u-net: a feasibility study for predicting dose distributions from
  contours using deep learning on prostate imrt patients, ArXiv Preprint
  ArXiv:1709.09233.

\bibitem{nguyen2018three}
D.~Nguyen, X.~Jia, D.~Sher, M.-H. Lin, Z.~Iqbal, H.~Liu, S.~Jiang,
  Three-dimensional radiotherapy dose prediction on head and neck cancer
  patients with a hierarchically densely connected u-net deep learning
  architecture, ArXiv Preprint ArXiv:1805.10397.

\bibitem{nguyen2019feasibility}
D.~Nguyen, T.~Long, X.~Jia, W.~Lu, X.~Gu, Z.~Iqbal, S.~Jiang, A feasibility
  study for predicting optimal radiation therapy dose distributions of prostate
  cancer patients from patient anatomy using deep learning, Scientific Reports
  9~(1) (2019) 1076.

\bibitem{nguyen20193d}
D.~Nguyen, X.~Jia, D.~Sher, M.-H. Lin, Z.~Iqbal, H.~Liu, S.~Jiang, 3d
  radiotherapy dose prediction on head and neck cancer patients with a
  hierarchically densely connected u-net deep learning architecture, Physics in
  Medicine \& Biology 64~(6) (2019) 065020.

\bibitem{burges1998tutorial}
C.~J. Burges, A tutorial on support vector machines for pattern recognition,
  Data Mining and Knowledge Discovery 2~(2) (1998) 121--167.

\bibitem{li2018improved}
H.~Li, J.~Weng, Y.~Shi, W.~Gu, Y.~Mao, Y.~Wang, W.~Liu, J.~Zhang, An improved
  deep learning approach for detection of thyroid papillary cancer in
  ultrasound images, Scientific Reports 8 (2018) 6600.

\bibitem{rao2018mitos}
S.~Rao, Mitos-rcnn: A novel approach to mitotic figure detection in breast
  cancer histopathology images using region based convolutional neural
  networks, ArXiv Preprint ArXiv:1807.01788.

\bibitem{cai2019efficient}
D.~Cai, X.~Sun, N.~Zhou, X.~Han, J.~Yao, Efficient mitosis detection in breast
  cancer histology images by rcnn, in: In Proceedings of the IEEE International
  Symposium on Biomedical Imaging, 2019, pp. 919--922.

\bibitem{akselrod2019cnn}
A.~Akselrod-Ballin, L.~Karlinsky, S.~Alpert, S.~Hashoul, R.~Ben-Ari, E.~Barkan,
  A cnn based method for automatic mass detection and classification in
  mammograms, Computer Methods in Biomechanics and Biomedical Engineering:
  Imaging \& Visualization 7~(3) (2019) 242--249.

\bibitem{ribli2018detecting}
D.~Ribli, A.~Horv{\'a}th, Z.~Unger, P.~Pollner, I.~Csabai, Detecting and
  classifying lesions in mammograms with deep learning, Scientific Reports
  8~(1) (2018) 4165.

\bibitem{redmon2016you}
J.~Redmon, S.~Divvala, R.~Girshick, A.~Farhadi, You only look once: Unified,
  real-time object detection, in: In Proceedings of the IEEE conference on
  computer vision and pattern recognition, 2016, pp. 779--788.

\bibitem{redmon2017yolo9000}
J.~Redmon, A.~Farhadi, Yolo9000: better, faster, stronger, in: In Proceedings
  of the IEEE conference on computer vision and pattern recognition, 2017, pp.
  7263--7271.

\bibitem{redmon2018yolov3}
J.~Redmon, A.~Farhadi, Yolov3: An incremental improvement, ArXiv Preprint
  ArXiv:1804.02767.

\bibitem{wu2019segmentation}
Y.~Wu, Z.~Cheng, Z.~Xu, W.~Wang, Segmentation is all you need, ArXiv Preprint
  ArXiv:1904.13300.

\bibitem{tran2018blood}
T.~Tran, O.-H. Kwon, K.-R. Kwon, S.-H. Lee, K.-W. Kang, Blood cell images
  segmentation using deep learning semantic segmentation, in: In Proceedings of
  the IEEE International Conference on Electronics and Communication
  Engineering, 2018, pp. 13--16.

\bibitem{doan2018diagnostic}
M.~Doan, I.~Vorobjev, P.~Rees, A.~Filby, O.~Wolkenhauer, A.~E. Goldfeld,
  J.~Lieberman, N.~Barteneva, A.~E. Carpenter, H.~Hennig, Diagnostic potential
  of imaging flow cytometry, Trends in Biotechnology 36~(7) (2018) 649--652.

\bibitem{chang2018cancer}
Y.~Chang, H.~Park, H.-J. Yang, S.~Lee, K.-Y. Lee, T.~S. Kim, J.~Jung, J.-M.
  Shin, Cancer drug response profile scan (cdrscan): a deep learning model that
  predicts drug effectiveness from cancer genomic signature, Scientific Reports
  8~(1) (2018) 8857.

\bibitem{xiao2018deep}
Y.~Xiao, J.~Wu, Z.~Lin, X.~Zhao, A deep learning-based multi-model ensemble
  method for cancer prediction, Computer Methods and Programs in Biomedicine
  153 (2018) 1--9.

\bibitem{badrinarayanan2017segnet}
V.~Badrinarayanan, A.~Kendall, R.~Cipolla, Segnet: A deep convolutional
  encoder-decoder architecture for image segmentation, IEEE transactions on
  pattern analysis and machine intelligence 39~(12) (2017) 2481--2495.

\bibitem{ardila2019end}
D.~Ardila, A.~P. Kiraly, S.~Bharadwaj, B.~Choi, J.~J. Reicher, L.~Peng, D.~Tse,
  M.~Etemadi, W.~Ye, G.~Corrado, et~al., End-to-end lung cancer screening with
  three-dimensional deep learning on low-dose chest computed tomography, Nature
  Medicine 25~(6) (2019) 954.

\bibitem{akselrod2016region}
A.~Akselrod-Ballin, L.~Karlinsky, S.~Alpert, S.~Hasoul, R.~Ben-Ari, E.~Barkan,
  A region based convolutional network for tumor detection and classification
  in breast mammography, in: Deep Learning and Data Labeling for Medical
  Applications, 2016, pp. 197--205.

\bibitem{drozdzal2016importance}
M.~Drozdzal, E.~Vorontsov, G.~Chartrand, S.~Kadoury, C.~Pal, The importance of
  skip connections in biomedical image segmentation, in: Deep Learning and Data
  Labeling for Medical Applications, 2016, pp. 179--187.

\bibitem{breiman2001random}
L.~Breiman, Random forests, Machine Learning 45~(1) (2001) 5--32.

\bibitem{klein2019maldi}
O.~Klein, F.~Kanter, H.~Kulbe, P.~Jank, C.~Denkert, G.~Nebrich, W.~D. Schmitt,
  Z.~Wu, C.~A. Kunze, J.~Sehouli, et~al., Maldi-imaging for classification of
  epithelial ovarian cancer histotypes from a tissue microarray using machine
  learning methods, Proteomics Clinical Applications 13~(1) (2019) 1700181.

\bibitem{coudray2018classification}
N.~Coudray, P.~S. Ocampo, T.~Sakellaropoulos, N.~Narula, M.~Snuderl,
  D.~Feny{\"o}, A.~L. Moreira, N.~Razavian, A.~Tsirigos, Classification and
  mutation prediction from non--small cell lung cancer histopathology images
  using deep learning, Nature Medicine 24~(10) (2018) 1559.

\bibitem{schaumberg2018h}
A.~J. Schaumberg, M.~A. Rubin, T.~J. Fuchs, H\&e-stained whole slide image deep
  learning predicts spop mutation state in prostate cancer, BioRxiv (2018)
  064279.

\bibitem{szegedy2016rethinking}
C.~Szegedy, V.~Vanhoucke, S.~Ioffe, J.~Shlens, Z.~Wojna, Rethinking the
  inception architecture for computer vision, in: In Proceedings of the IEEE
  conference on computer vision and pattern recognition, 2016, pp. 2818--2826.

\bibitem{chaudhary2018deep}
K.~Chaudhary, O.~B. Poirion, L.~Lu, L.~X. Garmire, Deep learning--based
  multi-omics integration robustly predicts survival in liver cancer, Clinical
  Cancer Research 24~(6) (2018) 1248--1259.

\bibitem{hartigan1979algorithm}
J.~A. Hartigan, M.~A. Wong, Algorithm as 136: A k-means clustering algorithm,
  Journal of the Royal Statistical Society. Series C (Applied Statistics)
  28~(1) (1979) 100--108.

\bibitem{devlin2018bert}
J.~Devlin, M.-W. Chang, K.~Lee, K.~Toutanova, Bert: Pre-training of deep
  bidirectional transformers for language understanding, ArXiv Preprint
  ArXiv:1810.04805.

\bibitem{dai2019transformer}
Z.~Dai, Z.~Yang, Y.~Yang, W.~W. Cohen, J.~Carbonell, Q.~V. Le,
  R.~Salakhutdinov, Transformer-xl: Attentive language models beyond a
  fixed-length context, ArXiv Preprint ArXiv:1901.02860.

\bibitem{yang2019xlnet}
Z.~Yang, Z.~Dai, Y.~Yang, J.~Carbonell, R.~Salakhutdinov, Q.~V. Le, Xlnet:
  Generalized autoregressive pretraining for language understanding, ArXiv
  Preprint ArXiv:1906.08237.

\bibitem{child2019generating}
R.~Child, S.~Gray, A.~Radford, I.~Sutskever, Generating long sequences with
  sparse transformers, ArXiv Preprint ArXiv:1904.10509.

\end{thebibliography}
\end{document}